\documentclass[12pt]{article}

\usepackage{amsmath}
\usepackage{amsfonts}
\usepackage{amssymb}
\usepackage{graphicx}%
\providecommand{\U}[1]{\protect\rule{.1in}{.1in}}

\usepackage{hyperref}

\begin{document}

\title{Classical and quantum integrable sigma models. \\ Ricci flow, ``nice duality'' and perturbed rational conformal field theories}

\author{Vladimir Fateev}
\date{}
\maketitle
\vspace{-1cm}

\begin{center}
\emph{Universit\'e Montpellier 2, Laboratoire Charles Coulomb,\\ UMR 5221, F-34095, Montpellier, France}
\end{center}

\begin{center}
\emph{Landau Institute for Theoretical Physics, 142432 Chernogolovka, Russia}
\end{center}

\vspace{0.1cm}

\begin{abstract}
We consider classical and quantum integrable sigma models and their relations with the solutions of renormalization group
equations. We say that an integrable sigma model possesses the ``nice'' duality property if the dual quantum field theory
has the weak coupling region. As an example, we consider the deformed $CP\left(  n-1\right)  $ sigma model
with additional quantum degrees of freedom. We formulate the dual integrable field theory and use perturbed conformal field
theory, perturbation theory, $S $-matrix, Bethe Ansatz and renormalization group methods to show that
this field theory has the ``nice'' duality property. We consider also an alternative approach to the analysis of sigma
models on the deformed symmetric spaces, based on the perturbed rational conformal field theories.

\end{abstract}

\newpage

\tableofcontents

\section{Introduction}

Duality plays an important role in the analysis of statistical, quantum field
and string theory systems. Usually it maps a weak coupling region of one
theory to the strong coupling region of the other and makes it possible to use
perturbative, semiclassical and renormalization group methods in different
regions of the coupling constant. For example, the well known duality between
Sine-Gordon and massive Thirring models \cite{C},\cite{M} together with
integrability plays an important role for the justification of exact
scattering matrix \cite{ZSG} in these theories. Another well known example of
the duality in two dimensional integrable systems is the weak-strong coupling
flow from affine Toda theories to the same theories with dual affine Lie
algebra \cite{AFT},\cite{COR},\cite{GR}. The phenomenon of electric-magnetic
duality in four dimensional $N=4$ supersymmetric gauge theories conjectured in
\cite{MO},\cite{NO} and developed for $N=2$ theories in \cite{SW} (and in many
subsequent papers) opens the possibility for the non-perturbative analysis of
the spectrum and phase structure in supersymmetric gauge field theories. The
remarkable field/string duality \cite{ADSCFT},\cite{MLD} leads to the
unification of the ideas and methods for the analysis of these seemingly
different quantum systems.

While known for many years the phenomenon of duality in quantum field theory
still looks rather mysterious and needs to be further analyzed. Such analysis
crucially simplifies for two-dimensional integrable relativistic systems.
These theories besides the Lagrangian formulation possess also an unambiguous
definition in terms of factorized scattering theory, which contains all
information about off- shell data of quantum theory. These data allow the use
of non-perturbative methods for the calculation of observables in integrable
field theories. The comparison of the observables calculated from the
scattering data and from the perturbative, semiclassical or renormalization
group analysis based on the Lagrangian formulation makes it possible in some
cases to justify the existence of two different (dual) Lagrangian
representations of a quantum theory.

The two particle factorized scattering matrix is a rather rigid object. It is
constrained by the global symmetries, factorization equation and unitarity and
crossing symmetry relations. After solving of these equations the scattering
matrix $S$ can contain one (or more) free parameter. At some value of this
parameter $\lambda=\lambda_{0}$ the scattering matrix $S\left(  \lambda
_{0}\right)  $ becomes the identity matrix and has a regular expansion around
this point. In many cases this expansion can be associated with the
perturbative expansion of some Lagrangian theory with parameter $b$ near some
free point. Sometimes there is a second point $\lambda$ $=$ $\lambda_{1}$
where $S\left(  \lambda\right)  $ reduces to identity matrix and admits a
regular expansion in $\left(  \lambda-\lambda_{1}\right)  $. If this expansion
can be associated with \ the perturbative expansion of another local
Lagrangian at small coupling $\gamma=\gamma(b)$, then the two different
Lagrangians describe the same theory, which has two different (dual)
perturbative regimes.

A more interesting situation occurs when $S\left(  \lambda\right)  $ has a
regular expansion in $\left(  \lambda-\lambda_{0}\right)  $ which agrees with
perturbative expansion in $b$ of some field theory with local action
$\mathcal{A}\left(  b\right)  $, but at the point $\lambda_{1}$ the $S$-matrix
tends to some \textquotedblleft rational\textquotedblright\ scattering matrix
corresponding to the $S$-matrix of a non-linear sigma model on a symmetric
space. Near the point $\lambda_{1}$ it can be considered as a deformation of a
symmetric scattering. In this case it is natural to search \ for the dual
theory as sigma model with target space looking as a deformed symmetric space.
The metric and other characteristics of sigma model on the manifold is subject
to very rigid conditions, namely non-linear renormalization group (RG)
equations \cite{FRI}. If one has found the solution of RG equations which
gives the observables in the sigma model theory, coinciding with those derived
from the factorized $S$- matrix theory one can conclude that field theory with
the action $\mathcal{A}\left(  b\right)  $ is dual to a sigma model on the
deformed symmetric space. The short distance behaviour of such theory can be
studied by RG and conformal field theory (CFT) methods. The agreement of the
CFT data, derived from the action $\mathcal{A}\left(  b\right)  $ (considered
as a perturbed CFT) with the data derived from RG for sigma model gives an
additional important test for the duality (\textquotedblright
nice\textquotedblright\ duality).

The analysis of integrable quantum SMs on the deformed symmetric spaces and
their dualities started in the papers \cite{FOZ},\cite{F3},\cite{F}. Later in
the papers \cite{CL},\cite{DD} the general classical SMs on the deformed
groups and cosets manifolds have been constructed.. Unfortunately, not all
these SMs, integrable classically, are integrable in quantum case. In
particular it happens for the cosets having $U\left(  1\right)  $ group in
denominator (see for example \cite{AB1}). In many cases this situation can be
improved by introduction of additional quantum degrees of freedom, which are
invisible in the classical limit.

We say that an integrable SM has the \textquotedblleft nice\textquotedblright%
\ duality if the dual integrable QFT has the weak coupling region. This
implies that we can study this theory by different methods (perturbation
theory, RG and CFT analysis) in different regimes. Note that the SMs with the
property of \textquotedblleft nice\textquotedblright\ duality form a very
small subspace in the space of all integrable quantum SMs on the deformed
symmetric spaces and such SMs with additional quantum degrees freedom.

A simple (but rather non-trivial) example is provided by the $CP\left(
n-1\right)  $ SM. This model is integrable classically but non-integrable (for
$n>2$) at the quantum level. After adding the massless fermion interacting
with $U\left(  1\right)  $ gauge field on $CP\left(  n-1\right)  $ (massless
axion, linearly coupled with the density of topological charge ) this SM
becomes integrable and its deformed version has the \textquotedblleft
nice\textquotedblright\ duality property. We study this theory in the main
part of this paper. Here we say a few words about non-integrable $CP\left(
n-1\right)  $ SMs.

These models were intensively studied during 70-80s due to their's similarity
with four-dimensional $SU(n)$ gauge QFTs. Namely, the $CP\left(  n-1\right)  $
SMs and $SU(n)$ gauge theories are asymptotically free, possess instantons,
and manifest the phenomenon of confinement.\footnote{For $n>2,$ both theories have
non-topological classical solutions, the role of which is not clear at the
moment.} 
It is natural that $CP\left(  n-1\right)  $ SMs served as
baby-laboratory for the analysis of $SU(n)$ gauge theory, in particular, for
analysis of instanton contributions \cite{FFS} and lattice simulations
\cite{IRV}.

The spectrum of the $CP\left(  n-1\right)  $ SMs in $\frac{1}{n}$ approach was
studied in \cite{DADA}. It was shown that besides the basic particles, which
are the only particles in the integrable version of this model, one has also
the particles which are their bound states, confined by Coulomb forces. The
addition of fermion (axion) produce an essential restructure of the spectrum.
Of course, the influence of the axion on the spectrum of gauge theory is a
more interesting and much more complicated problem.

The spectrum of the deformed non-integrable $CP\left(  n-1\right)  $ SMs seems
to be qualitatively the same as that of the undeformed models and can not be
studied by perturbative methods. Due to the \textquotedblleft
nice\textquotedblright\ duality, in the integrable $CP\left(  n-1\right)  $
SMs with axion there is a weak coupling region. In this region the basic
particles also form the bound state, which disappear from the spectrum outside
the perturbative region. It is possible however that in a non-integrable QFT
they survive in the strong coupling (SM) regime. We hope to return to this
problem in a future publication.

This paper is organized as follows. In section 2 we describe the basic CFTs,
which can be formulated in terms of $2n-1$ bosonic fields, and their primary
fields are the exponents of these fields. We calculate the reflection
amplitudes in these CFTs which are important for the calculation of UV
asymptotics in perturbed CFTs. These amplitudes serve also for identification
of CFTs in different representations. In particular, for justification of dual
SM representations.

In section 3 we explain the general properties of deformed $CP\left(
n-1\right)  $ SMs with fermion and write the action of perturbed CFTs,
constructed in section 2. We conjecture that these QFTs provide a dual
description of deformed $CP\left(  n-1\right)  $ SMs with fermion.

In section 4 we represent the action of dual QFT in the form suitable for the
perturbation theory in parameter $b.$ We provide non-local integrals of motion
which form the Borel subalgebra of $SU(n)_{\mathrm{q}}$ and generate
$SU(n)_{\mathrm{q}}$ symmetry of the scattering theory. We describe the
spectrum and scattering theory of this QFT.

In section 5 we use the Bethe Ansatz approach to derive the exact relations
between the parameters of action and scattering theory in this QFT. We
calculate the observables, which can be compared with the observables
calculated using the dual SM description of our QFT.

In section 6 we consider classical and quantum integrable SMs on the deformed
symmetric spaces. We discuss Ricci flows in these SMs and the relation between
the parameters of integrable SMs with the parameters of their scattering
theory. We calculate the observables in integrable deformed $CP\left(
n-1\right)  $ SMs and show that in the scaling (one loop) approximation they
coincide with the observables calculated in the Bethe Ansatz approach.

In section 7 we consider the conformal limit of the deformed $CP\left(
n-1\right)  $ SMs. For simplicity of equations we consider the case $n=3.$ We
calculate the reflection amplitudes associated with this conformal SM and show
that they coincide with reflection amplitudes calculated in section 2 i.e.
these CFTs are dual. We use these amplitudes to get the UV asymptotics of
effective central charge in the deformed $CP\left(  n-1\right)  $ SMs on the
circle of length $R.$

In section 8 we study the second integrable perturbation of CFTs described in
section 2. We study the scattering theory of these QFTs and see that at small
and large values of the coupling constant $b$ they can be studied by
perturbation theory in $b$ and $\frac{1}{b}$ i.e. it has two different dual
representation. In the strong coupling regime they can be described by the
action with the SM part coinciding with the conformal limit of the deformed
$CP\left(  n-1\right)  $ SMs and the potential part described by tachyon.
After a simple analytical continuation in the coupling constant the SM part of
the actions become singular. These action describe integrable classical models
but after the quantization these QFTs are well defined only for discrete
values of coupling constant.

In section 9 we discuss the rational CFTs, which are closely related
with conformal SMs with singular actions and the discrete values of coupling
constant. We show that these CFTs perturbed by proper fields describe
non-integrable deformed $CP\left(  n-1\right)  $ SMs with topological
parameter and integrable deformed $CP\left(  n-1\right)  $ with fermion
(axion). We consider more general CFTs represented by the cosets $\frac
{G_{m}\times G_{l}}{G_{m+l}}$ and study their deformations by different fields
in different regions of integers $m,l$ and $h$ (Coxeter number of $G$). We
study their RG propertied and show that these QFTs provide an independent
description of a large variety of SMs on the deformed symmetric spaces. Finally
we give a simple inequality which provides a necessary condition for a sigma
model to have \textquotedblleft nice\textquotedblright\ duality.

\section{Conformal field theory and reflection amplitudes }

We consider \ CFT, which has the hidden $sl(n)$and $sl(n-1)-$ symmetries and
can be described by $2n-1$ fields%
\begin{equation}
\{\varphi,\vartheta,\vartheta_{0},\vartheta_{n}\}=\{(\varphi_{1}%
,..,\varphi_{n-1}),(\vartheta_{1},..,\vartheta_{n-2}),\vartheta_{0}%
,\vartheta_{n}\} \label{f}%
\end{equation}
We introduce the parameters $a,b$ satisfying the relation%
\begin{equation}
a^{2}-b^{2}=1 \label{ab}%
\end{equation}
and denote as $c$ the parameter%
\begin{equation}
c=\frac{\sqrt{n+b^{2}}}{\sqrt{n(n-1}} \label{cc}%
\end{equation}
Let $h_{j}$ are $n$ vectors in $n-1$ dimensional space (the wights of
fundamental representation of $sl(n)$) with the scalar products:
\[
h_{i}\cdot h_{j}=\delta_{i,j}-\frac{1}{n}%
\]
and $\eta_{j}$ the vectors in $n-2$ dimensional space (the wights of
fundamental representation of $sl(n-1),\eta_{i}\cdot\eta_{j}=\delta
_{i,j}-\frac{1}{n-1}$). Then $2(n-1)$ \textquotedblleft screening
charges\textquotedblright\ can be written in the form of $(n-1)$ pairs of
fields $\{V_{1,}V_{-2}\},\{V_{2,}V_{-3}\},..,\{V_{n-1,}V_{-n}\},$ where
$\{V_{k,}V_{-k-1}\}$ ($k=1,..,n-1$) is:
\begin{equation}
\{\mu\exp(b(h_{k}\cdot\varphi)+\mathrm{i}a(\eta_{k}\cdot\vartheta
)+\mathrm{i}c\vartheta_{0}),\mu\exp(-b(h_{k+1}\cdot\varphi)-\mathrm{i}%
a(\eta_{k}\cdot\vartheta)-\mathrm{i}c\vartheta_{0})\} \label{cpai}%
\end{equation}
We introduce $n-1$ and $n-2$ dimensional vectors $\rho$ and $\rho_{1},$ which
are the halves of sum of the roots of $sl(n),$ and $sl(n-1)$ and the dilaton
field%
\begin{equation}
\Phi=\frac{1}{b}(\rho\cdot\varphi)+\frac{\mathrm{i}}{a}(\rho_{1}\cdot
\vartheta)=(Q\cdot\varphi)+\mathrm{i}(Q_{1}\cdot\vartheta). \label{d1}%
\end{equation}
Then the theory described by the action $\mathcal{A}_{CFT}=\mathcal{A}%
_{0}+\mathcal{A}_{C},$ where $\mathcal{A}_{0}=\int d^{2}x\frac{1}{8\pi
}(\partial_{\mu}\vartheta_{n}\partial_{\mu}\vartheta_{n})$ is the action of a
free field $\vartheta_{n}$ and%
\begin{equation}
\mathcal{A}_{C}=\int d^{2}x\left[\frac{1}{8\pi}\bigl((\partial_{\mu}\varphi\cdot
\partial_{\mu}\varphi)+(\partial_{\mu}\vartheta\cdot\partial_{\mu}%
\vartheta)+(\partial_{\mu}\vartheta_{0}\partial_{\mu}\vartheta_{0})\bigr)
+U(\varphi,\vartheta,\vartheta_{0})+R_{2}\Phi\right], \label{ac1}%
\end{equation}
with $U(\varphi,\phi,\phi_{0})=\mu\sum_{j=1}^{n-1}\left(  e^{b(h_{j}%
\cdot\varphi)+\mathrm{i}a(\eta_{j}\cdot\vartheta)+\mathrm{i}c\vartheta_{0}%
}+e^{-b(h_{j+1}\cdot\varphi)-\mathrm{i}a(\eta_{j}\cdot\vartheta)-\mathrm{i}%
c\vartheta_{0}}\right)  ,$ defines the CFT with central charge:
\begin{equation}
c=1+(n-1)\left(  2+\frac{n(n+1)}{b^{2}}-\frac{n(n-2)}{a^{2}}\right)  .
\label{cn}%
\end{equation}
As we see later, the CFT (\ref{ac1}) describes the dual representation for coset CFT
$\frac{SL\left(  n\right)  }{SL\left(  n-1\right)  U\left(  1\right)  }.$ For
$n=2$ it coincides with Sine-Liouville CFT (see for example \cite{FS})

The primary fields in this CFT (we do not consider here the free field
$\vartheta_{n}$) are the exponential fields:%
\[
V_{A,B,C}=\exp\left(  (A\cdot\varphi)+\mathrm{i}(B\cdot\phi)+\mathrm{i}%
C\vartheta_{0}\right)
\]
with dimension
\[
\Delta(A,B,C)=\frac{1}{2}\left(  -A^{2}+2(A\cdot Q)+B^{2}-2(B\cdot
Q_{1})+C^{2}\right)  .
\]
In the space of the fields $V_{A,B,C}$ acts the Weyl group $\mathbf{W,}$ which
is the product of groups $\mathbf{w}_{n}$ and $\mathbf{w}_{n-1}$ of the Lie
algebras $sl(n)$ and $sl(n-1).$ Here we consider the group $\mathbf{w}_{n},$
which acts as $V_{A,B,C}\rightarrow\mathbf{R}_{\widehat{s}}V_{Q+\widehat
{s}(A-Q),B,C}$ $(\widehat{s}\subset\mathbf{w}_{n})$ where $\mathbf{R}%
_{\widehat{s}}$ is the reflection amplitude \cite{ZZ,FR}.\footnote{Reflection amplitudes 
identify the CFTs which admit the free field representation.} 
We normalize our fields by the condition:
\[
\langle V_{A,B,C}(x)V_{2Q-A,2Q_{1}-B,-C}(0)\rangle=|x|^{-4\Delta(A,B,C)}%
\]
Then the reflection amplitude can be expressed through two point function:
\begin{equation}
\left\langle V_{Q+\mathbf{a},Q_{1}+\mathbf{b},C}(x)\right\rangle
V_{Q+\mathbf{a}^{\ast},Q_{1}-\mathbf{b},-C}(0)\rangle=\mathbf{R}%
(\mathbf{a},\mathbf{b},C)|x|^{-4\Delta} \label{rr}%
\end{equation}
where vectors $\mathbf{a}^{\ast}$ and $\mathbf{a}$ are related by the usual
conjugation: $(\mathbf{a}\cdot e_{j})=(\mathbf{a}^{\ast}\cdot e_{n-j}%
),(\mathbf{a}\cdot h_{j})=-(\mathbf{a}^{\ast}\cdot h_{n+1-j}),$ here
$e_{j}=h_{j}-h_{j+1},$ are the simple roots of $sl(n).$

The function $\mathbf{R}(\mathbf{a},\mathbf{b},C)$ in the CFT (\ref{ac1}) can
be calculated exactly using the integral relation \cite{BF},\cite{F1L}.
\begin{align}
\int\mathcal{D}_{n}\left(  x\right)
{\textstyle\prod\limits_{i=1}^{n}}
{\textstyle\prod\limits_{j=1}^{n+m+2}}
|x_{i}-t_{j}|^{2p_{j}}d^{2n}x  &  =%
{\textstyle\prod\limits_{j=1}^{n+m+2}}
\frac{\Gamma\left(  1+p_{j}\right)  }{\Gamma\left(  -p_{j}\right)  }%
{\textstyle\prod\limits_{k<J}}
|t_{k}-t_{j}|^{2p_{k}+2p_{j}+2}\nonumber\\
&  \times\int\mathcal{D}_{m}\left(  y\right)
{\textstyle\prod\limits_{i=1}^{m}}
{\textstyle\prod\limits_{j=1}^{n+m+2}}
|y_{i}-t_{j}|^{-2-2p_{j}}d^{2m}y,
\end{align}
where $\mathcal{D}_{n}\left(  x\right)  =$ $%
{\textstyle\prod\limits_{i<j}}
|x_{i}-x_{j}|^{2},$ $d^{2n}x=\frac{1}{n!\pi^{n}}%
{\textstyle\prod\limits_{i=1}^{n}}
d^{2}x_{i},\
{\textstyle\sum\limits_{j=1}^{n+m+2}}
p_{j}=-n-1;$ and has the form:
\begin{align}
\mathbf{R}(\mathbf{a},\mathbf{b},C)  &  =%
{\displaystyle\prod\limits_{j=1}^{[n/2]}}
\left(  \frac{\Gamma(\frac{1}{2}-b(\mathbf{a}\cdot h_{j})\mathbf{+}%
a(\mathbf{b}\cdot\eta_{j})+cC)\Gamma(\frac{1}{2}-b(\mathbf{a}^{\ast}\cdot
h_{j})\mathbf{-}a(\mathbf{b}\cdot\eta_{j})-cC)}{\Gamma(\frac{1}{2}%
+b(\mathbf{a}\cdot h_{j})\mathbf{-}a(\mathbf{b}\cdot\eta_{j})-cC)\Gamma
(\frac{1}{2}+b(\mathbf{a}^{\ast}\cdot h_{j})\mathbf{+}a(\mathbf{b}\cdot
\eta_{j})+cC)}\right)  ^{n-2j+1}\nonumber\\
&  \times\left(  \frac{\pi\mu}{b^{2}}\right)  ^{2(\mathbf{a}\cdot\rho)/b}%
{\displaystyle\prod\limits_{\alpha>0}}
\frac{\Gamma(1+b(\mathbf{a}\cdot e_{\alpha}))\Gamma(1+\frac{1}{b}%
(\mathbf{a}\cdot e_{\alpha}))}{\Gamma(1-b(\mathbf{a}\cdot e_{\alpha}%
))\Gamma(1-\frac{1}{b}(\mathbf{a}\cdot e_{\alpha}))} \label{ra1}%
\end{align}
where the last product runs over all positive roots. This function corresponds
to the maximal reflection $\mathbf{a}\rightarrow-$ $\mathbf{a}^{\ast}$ and can
be represented in the form:%
\begin{equation}
\mathbf{R}(\mathbf{a},\mathbf{b},C)=\frac{\mathbf{A}(-\mathbf{a}^{\ast
},\mathbf{b},C)}{\mathbf{A}(\mathbf{a},\mathbf{b},C)}%
\end{equation}
where $\frac{1}{\mathbf{A}(\mathbf{a},\mathbf{b},C)}~$is entire function,
which includes all products in the denominator of (\ref{ra1}).
\begin{align}
\mathbf{A}(\mathbf{a},\mathbf{b},C)  &  =%
{\displaystyle\prod\limits_{j=1}^{[n/2]}}
\left(  \Gamma(\frac{1}{2}+b(\mathbf{a}\cdot h_{j})\mathbf{-}a(\mathbf{b}%
\cdot\eta_{j})-cC)\Gamma(\frac{1}{2}+b(\mathbf{a}^{\ast}\cdot h_{j}%
)\mathbf{+}a(\mathbf{b}\cdot\eta_{j})+cC))\right)  ^{n-2j+1}\nonumber\\
&  \times\left(  \frac{\pi\mu}{b^{2}}\right)  ^{-(\mathbf{a}\cdot\rho)/b}%
{\displaystyle\prod\limits_{\alpha>0}}
\Gamma(1-b(\mathbf{a}\cdot e_{\alpha}))\Gamma(1-\frac{1}{b}(\mathbf{a}\cdot
e_{\alpha})). \label{qam}%
\end{align}

The general reflection amplitudes $\mathbf{R}_{\widehat{s}}(\mathbf{a}%
,\mathbf{b},C)$ are:%
\begin{equation}
\mathbf{R}_{\widehat{s}}(\mathbf{a},\mathbf{b},C)=\frac{\mathbf{A}(\widehat
{s}\mathbf{a},\mathbf{b},C)}{\mathbf{A}(\mathbf{a},\mathbf{b},C)}. \label{Rs}%
\end{equation}
These functions play the important role in the analysis of UV asymptotics in
perturbed CFTs and in calculation of expectation values of the fields in these
theories. In this paper we also use these functions in section 7 to conjecture
the dual SM representation for CFT (\ref{ac1}).

\section{Deformed $CP(n-1)$ models with fermion and dual QFT}

The $CP(n-1)$ models are two-dimensional asymptotically free sigma-models on
$(n-1)$ complex dimensional K\"{a}hler space with an $SU(n)$ invariant metric.
As all sigma models with K\"{a}hler metric they possess instantons which are
the holomorphic functions of world -sheet coordinates . The K\"{a}hler metric
can be written as $ds^{2}=$ $g_{ab^{\ast}}dw_{a}dw_{b^{\ast}}^{\ast}$ and
defines a two form $\kappa=$\textrm{i}$g_{ab^{\ast}}dw_{a}\wedge dw_{b^{\ast}%
}^{\ast}$ This form after proper normalization defines the density of
topological charge and is closed (but not exact). It means that locally
$\kappa=d\beta,$ where $\beta$ is a one form. \ In the world sheet coordinates
this form defines the $U(1)$ gauge field $B$ which can interact with a complex
fermion field $\chi.$ We call the theory with action:
\begin{equation}
\mathcal{A}_{d}=\mathcal{A}_{sm}+\int\frac{d^{2}x}{4\pi}\overline{\chi
}\mathrm{i}\gamma_{\mu}(\partial_{\mu}+\mathrm{i}B_{\mu})\chi\label{ad}%
\end{equation}
where $\mathcal{A}_{sm}$ is the action of sigma model, as deformed $CP(n-1)$
models with fermion. After the deformation the $SU(n)$ symmetry of the metric
is broken up to $U(1)^{n-1}.$ One more $U(1)$ symmetry comes from the fermion
interacting with $B,$ so the full symmetry of theory is $U(1)^{n}.$ We note
that the introduction of fermionic field interacting with sigma model cancels
the contributions of the fields with a non-zero topological charge and long
range interaction in the QFT (\ref{ad}).

To construct the integrable QFT which can be considered as perturbed CFT
(\ref{ac1}) we have two possibilities. To add the affine pair of fields, or to
add the term $\mu_{1}e^{e_{0}\cdot\varphi},$ where $e_{0}$ is the affine root
of $sl(n).$

To discuss the relation with sigma model (\ref{ad}) we consider the first
possibility. Namely, we add the affine pair (which we denote as $\{V_{n,}%
V_{-n-1}\}$). It is convenient to make the orthogonal transformation from $n$
the fields $\{(\vartheta_{1},..,\vartheta_{n-2}),\vartheta_{0},\vartheta
_{n}\}$ \ to $n$ fields $\{(\phi_{1},..,\phi_{n-1}),\phi_{0}\}$ in such way
that pairs $\{V_{k,}V_{-k-1}\}$ ($k=1,..,n-1$) (\ref{cpai}) transform to
\[
\{\mu\exp(b(h_{k}\cdot\varphi)+\mathrm{i}a(h_{k}\cdot\phi)+\mathrm{i}%
n^{-1/2}\phi_{0}),\mu\exp(-b(h_{-k-1}\cdot\varphi)-\mathrm{i}a(h_{k}\cdot
\phi)-\mathrm{i}n^{-1/2}\phi_{0})\}
\]
and the affine pair $\{V_{n,}V_{-n-1}\}$ is:
\[
\{\mu\exp(b(h_{n}\cdot\varphi)+\mathrm{i}a(h_{n}\cdot\phi)+\mathrm{i}%
n^{-1/2}\phi_{0}),\mu\exp(-b(h_{1}\cdot\varphi)-\mathrm{i}a(h_{n}\cdot
\phi)-\mathrm{i}n^{-1/2}\phi_{0})\}
\]

The action of our QFT will now have a form:
\begin{equation}
\mathcal{A}_{sl(n)}^{(1)}=\int d^{2}x\left[  \frac{1}{8\pi}((\partial_{\mu
}\varphi\cdot\partial_{\mu}\varphi)+(\partial_{\mu}\phi\cdot\partial_{\mu}%
\phi)+(\partial_{\mu}\phi_{0}\partial_{\mu}\phi_{0}))+\mathbf{U}(\varphi
,\phi,\phi_{0})\right]  \label{asl}%
\end{equation}
where $\mathbf{U}(\varphi,\phi,\phi_{0})$ can be written in the symmetric form
(here and latter $h_{n+1}=h_{1}$)\footnote{The action (\ref{asl}), (\ref{Aexp}) coincides with the action of $\mathfrak{gl}%
(n|n)$\ Toda model \cite{LIT} after a simple transformation of fields and
exclusion of one of them.}
\begin{equation}
\mathbf{U}(\varphi,\phi,\phi_{0})=%
{\displaystyle\sum\limits_{j=1}^{n}}
\mu\left(  e^{b(h_{j}\cdot\varphi)+\mathrm{i}a(h_{j}\cdot\phi)+\mathrm{i}%
n^{-1/2}\phi_{0}}+e^{-b(h_{j+1}\cdot\varphi)-\mathrm{i}a(h_{j}\cdot
\phi)-\mathrm{i}n^{-1/2}\phi_{0}}\right)  \label{Aexp}%
\end{equation}
Besides the discreet symmetries related to permutations of $h_{i}$ the action
(\ref{asl}) possesses the $U(1)^{n}$ symmetry associated with the following
transformations:
\[
\phi\rightarrow\phi+\frac{2\pi}{a}\sum_{i=1}^{n-1}e_{i}m_{i},\qquad
\phi_{0}\rightarrow\phi_{0}+2\pi n^{1/2}m_{0},
\qquad
(m_{i},m_{0})\subset Z
\]
In the next section we represent this action in the form suitable for
perturbation theory in $b$. In this form the $U(1)^{n}$-symmetry will be obvious.

\section{Action of dual QFT admitting perturbative expansion, non-local
integrals and scattering theory}

Let us consider the case $n=2,$ which is slightly special. Here $h_{1}%
=\frac{1}{\sqrt{2}},h_{2}=-\frac{1}{\sqrt{2}}$ and the action has the form:
\begin{equation}
\mathcal{A}_{sl(2)}^{(1)}=\int d^{2}x\left[  \frac{1}{8\pi}(\partial_{\mu
}\varphi\partial_{\mu}\varphi+\partial_{\mu}\phi\cdot\partial_{\mu}%
\phi+\partial_{\mu}\phi_{0}\partial_{\mu}\phi_{0})+\mathbf{U}(\varphi
,\phi,\phi_{0})\right]  ,\label{sl2}%
\end{equation}
where $\mathbf{U}=2\mu\left(  e^{b\varphi/\sqrt{2}}\cos\left(  \frac
{(a\phi+\phi_{0})}{\sqrt{2}}\right)  +e^{-b\varphi/\sqrt{2}}\cos\left(
\frac{(-a\phi+\phi_{0})}{\sqrt{2}}\right)  \right)  .$ This theory belongs to
two parameter family of integrable QFTs \cite{F}, with $a_{1}^{2}+a_{2}%
^{2}-b^{2}=1$ and
\begin{equation}
\mathbf{U}_{2p}=2\mu\left(  e^{b\varphi/\sqrt{2}}\cos\left(  \frac{(a_{1}%
\phi+a_{2}\phi_{0})}{\sqrt{2}}\right)  +e^{-b\varphi/\sqrt{2}}\cos\left(
\frac{(-a_{1}\phi+a_{2}\phi_{0})}{\sqrt{2}}\right)  \right)  .\label{2p}%
\end{equation}
Besides the local IM, this theory possesses non-local ones generated by chiral
fields
\begin{align}
J_{1}^{(\pm)} &  =e^{\pm\mathrm{i}\sqrt{2}/a_{1}\phi}%
(\pm\mathrm{i}a_{1}\partial\phi\pm\mathrm{i}a_{2}\partial\phi_{0}%
+b\partial\varphi),
\quad 
J_{2}^{(\pm)}=e^{\pm\mathrm{i}\sqrt{2}/a_{2}\phi_{0}}
(\pm\mathrm{i}a_{1}\partial\phi\pm\mathrm{i}a_{2}%
\partial\phi_{0}+b\partial\varphi),\nonumber\\
I_{1} &  =e^{\sqrt{2}/b\varphi}(\mathrm{i}a_{1}\partial\phi+\mathrm{i}%
a_{2}\partial\phi_{0}+b\partial\varphi),
\quad 
I_{2}=e^{-\sqrt{2}/b\varphi
}(\mathrm{i}a_{1}\partial\phi-\mathrm{i}a_{2}\partial\phi_{0}-b\partial
\varphi)\label{int2p}%
\end{align}
Non-local integrals of motion generated by currents $J_{1}^{(\pm)},J_{2}^{(\pm)}$ form the
quantum group symmetry $SU(2)_{\mathrm{q}_{1}}\otimes SU_{\mathrm{q}_{2}}(2)$
($\mathrm{q}_{j}=\exp\left(  \frac{2\pi\mathrm{i}}{a_{j}^{2}}\right)  $) and
it is not surprising that\ scattering matrix of this QFT can be expressed in
terms of a direct product of two $S$-matrices of Sine-Gordon model. If we
denote the coupling constant in SG model as $\beta_{SG}^{2}=\frac{p}{p+1},$
the scattering matrix of the theory (\ref{2p}) will be
\begin{equation}
S_{2p}(\theta)=-S_{p_{1}}^{(SG)}(\theta)\otimes S_{p_{2}}^{(SG)}%
(\theta)\label{SSp}%
\end{equation}
here $\theta$ $=$ $\theta_{1}-\theta_{2}$ is the relative rapidity of
colliding particles and $S_{p_{i}}^{(SG)}(\theta)$ are $S$-matrices of
SG-models with $\beta_{i,SG}^{2}=\frac{p_{i}}{p_{i}+1}$ and $p_{i}=a_{i}^{2}.$
This fact shows that the deformed $CP(1)-$model with fermion possesses
two-parameter family of deformation. In our case $p_{1}=a_{1}^{2}=1+b^{2}%
,$\ $p_{2}=1,$ and $S$-matrix of the QFT (\ref{sl2}) will be the matrix
\begin{equation}
S=-S_{a^{2}}^{(SG)}(\theta)\otimes S_{1}^{(SG)}(\theta)=-S_{p_{1}}%
^{(SG)}(\theta)\otimes S_{1}^{(SG)}(\theta)\label{1p1}%
\end{equation}
$S$-matrix $S_{1}^{(SG)}(\theta)$ coincides with $S$-matrix of free fermion
theory, but the full scattering matrix is not trivial. We will not discuss in
this section the two-parametric deformations of action $\mathcal{A}%
_{sl(2)}^{(1)}$ but to understand the structure of the particles and the
perturbation theory for small $b,$ it is convenient to start with the case
$n=2.$ In this case after application of Coleman-Mandelstam $2d$
correspondence \cite{C},\cite{M} between fermions and bosons
\begin{equation}
\frac{1}{2}\partial_{\mu}\xi\partial_{\mu}\xi\rightarrow\mathrm{i}%
\overline{\psi}\gamma_{\mu}\partial_{\mu}\psi,\quad\partial_{\mu}%
\xi\rightarrow\overline{\psi}\gamma_{\mu}\psi,\quad e^{\pm\mathrm{i}\xi
}\rightarrow\overline{\psi}(1+\gamma_{5})\psi\label{CM}%
\end{equation}
we can rewrite the action $\mathcal{A}_{sl(2)}^{(1)}$ (after transformation
$\xi_{1}=\frac{1}{\sqrt{2}}(a\phi+\phi_{0}),$ $\xi_{2}=\frac{1}{\sqrt{2}%
}(-a\phi+\phi_{0})$) as
\[
\mathcal{A}_{sl(2)}^{(1)}=\int(L_{F}+L_{FB}+L_{B})d^{2}x
\]
where%
\begin{align}
L_{F} &  =\frac{1}{4\pi}\left(
{\displaystyle\sum\limits_{i=1}^{2}}
(\mathrm{i}\overline{\psi}_{i}\gamma_{\mu}\partial_{\mu}\psi_{i}-\frac{b^{2}%
}{2(1+b^{2})}(\overline{\psi}_{i}\gamma_{\mu}\psi_{i})^{2})+\frac{b^{2}}
{(1+b^{2})}(\overline{\psi}_{1}\gamma_{\mu}\psi_{1})(\overline{\psi}_{2}\gamma_{\mu}\psi_{2})\right)  ,
\nonumber\\
L_{FB} &  =\frac{M_{0}}{4\pi}\left(  e^{b\frac{\varphi}{\sqrt{2}}}%
\overline{\psi}_{1}\psi_{1}+e^{-b\frac{\varphi}{\sqrt{2}}}\overline{\psi}_{2}\psi_{2}\right)  ,
\qquad 
L_{B}=\frac{\partial_{\mu}\varphi\partial_{\mu}\varphi}{8\pi}
+\frac{M_{0}^{2}}{4\pi b^{2}}\left(  e^{\sqrt{2}b\varphi}+e^{-\sqrt{2}b\varphi}\right)  .
\label{su2}%
\end{align}
The last term in $L_{B}$ is the usual contact counterterm which cancels the
divergencies, coming from fermion loops. The particles in this QFT are
fermions $\psi_{1},\psi_{1}^{\ast};\psi_{2},\psi_{2}^{\ast},$ with
$U(1)\otimes U(1)$ charges $(1,-1)\otimes(1,-1).$ It is easy to check that the
perturbation theory in $b$ coincides with small $\ b$ expansion of $S$-matrix
of our theory.

In general case the QFT (\ref{asl}) possesses non-local IM generated by the
currents $J_{i},I_{i}\quad(i=1,..,n):$%
\begin{align*}
J_{i}  &  =\exp\left(  \frac{\mathrm{i}}{a}(e_{i}\cdot\phi)\right)
(\mathrm{i}a(h_{i}\cdot\partial\phi)+b(h_{i+1}\cdot\partial\varphi
)+\mathrm{i}\partial\phi_{0}n^{-1/2}),\\
I_{i}  &  =\exp\left(  \frac{1}{b}(e_{\mathrm{i}}\cdot\varphi)\right)
(\mathrm{i}a(h_{i}\cdot\partial\phi)+b(h_{i}\cdot\partial\varphi
)+\mathrm{i}\partial\phi_{0}n^{-1/2})
\end{align*}
where $e_{n}=h_{n}-h_{1}$ and $h_{n+1}=h_{1}.$ In appendix $A$ we provide the
examples of local integrals of motion.

The IM corresponding to currents $J_{i}$ generate $SU(n)_{\mathrm{q}}$ (with
$\mathrm{q}=\exp\left(  \frac{2\pi\mathrm{i}}{a^{2}}\right)  )\ $symmetry of
the scattering theory of this QFT. The solution of Yang-Baxter equations with
this symmetry was constructed by Cherednik \cite{Ch}. To understand better the
content of particles and spectrum of this theory, it is useful to use again
the Coleman-Mandelstam $2d$ correspondence \cite{C},\cite{M} between fermion
and bosons (\ref{CM}). We introduce the fields
\[
\xi_{i}=a(h_{i}\cdot\phi)+\phi_{0}n^{-1/2}%
\]
and projectors $\gamma_{+},\gamma_{-}:\gamma_{\pm}=\frac{1}{2}(1\pm\gamma
_{5}).$ Then the action (\ref{asl}) after application of Coleman-Mandelstam
rules can be rewritten in the form:
\begin{equation}
\mathcal{A}_{sl(n)}^{(1)}=\int d^{2}x\left(  L_{F}+L_{B}+L_{FB}\right)  ,
\label{afb}%
\end{equation}
where:
\begin{align}
L_{F}  &  =\frac{1}{4\pi}\left(
{\displaystyle\sum\limits_{i=1}^{n}}
(\mathrm{i}\overline{\psi}_{i}\gamma_{\mu}\partial_{\mu}\psi_{i}%
-g_{1}(\overline{\psi}_{i}\gamma_{\mu}\psi_{i})^{2})+g_{2}%
{\displaystyle\sum\limits_{i\neq j}^{n}}
(\overline{\psi}_{i}\gamma_{\mu}\psi_{i})(\overline{\psi}_{j}\gamma_{\mu}%
\psi_{j})\right)  ,\nonumber\\
L_{FB}  &  =\frac{M_{0}}{4\pi}%
{\displaystyle\sum\limits_{j=1}^{n}}
\left(  e^{b(h_{j}\cdot\varphi)}\overline{\psi}_{i}\gamma_{+}\psi
_{i}+e^{-b(h_{j}+_{1}\cdot\varphi)}\overline{\psi}_{i}\gamma_{-}\psi
_{i},\right)  ,\nonumber\\
L_{B}  &  =\frac{1}{8\pi}\left(  (\partial_{\mu}\varphi\cdot\partial_{\mu
}\varphi)+\frac{2M_{0}^{2}}{b^{2}}%
{\displaystyle\sum\limits_{j=0}^{n-1}}
e^{b(e_{j}\cdot\varphi)}\right),\quad
g_{1}=\frac{(n-1)b^{2}}{n(1+b^{2})},
\quad
g_{2}=\frac{b^{2}}{n(1+b^{2})}.
\label{acf}%
\end{align}
The last term in $L_{B}$ , which has a form of $sl_{n}$ Toda potential
($e_{0}$ is the affine root of $sl_{n}$ ), is the usual contact counterterm
which cancels the divergencies, coming from fermion loops. This QFT has a
broken $\mathbf{P}$ symmetry $\ $but possesses $\mathbf{PT}$ and $\mathbf{C}$
symmetries. For small $b$ it has $2n$ particles $\psi_{i},\psi_{i}^{\ast}$
with mass $M,$ and $(n-1)$ scalar particles $\varphi_{i}$ which can be
considered as bound states of Fermi particles with masses $M_{j}=2M\sin\left(
\frac{\pi j}{n}\right)  +O(b^{2}).$ As we see later, for finite $b$ they
disappear from spectrum.

The scattering theory of basic particles $\psi_{i}\rightarrow i,,\psi
_{i}^{\ast}\rightarrow\overline{i}$ has $U\left(  1\right)  ^{n}$ symmetry.
For $n>2$ the $S$-matrix of these particles is also restricted by
$SU(n)_{\mathrm{q}}$ symmetry and has no backward scattering amplitudes of
particles and antiparticles. It also should satisfy $\mathbf{PT}$,
$\mathbf{C,}$ and crossing symmetries:%
\[
S_{j_{1}j_{2}}^{i_{1}i_{2}}(\theta)=S_{i_{2}i_{1}}^{j_{2}j_{1}}(\theta),
\qquad
S_{j_{1}j_{2}}^{i_{1}i_{2}}(\theta)=S_{\overline{j}_{1}\overline{j}_{2}%
}^{\overline{i}_{1}\overline{i}_{2}}(\theta),\qquad S_{j_{1}j_{2}}^{i_{1}i_{2}%
}(\theta)=S_{j_{2}\overline{i}_{2}}^{i_{2}\overline{j}_{2}}(\mathrm{i}%
\pi-\theta).
\]
It is convenient to write these amplitudes in the form:%
\begin{align}
S_{ii}^{ii}(\theta)  &  =-F(\theta),\quad S_{ij}^{ij}(\theta)=-F(\theta
)\frac{\sinh(\frac{n}{2}\lambda\theta)}{\sinh(\frac{n}{2}\lambda(\theta
-i\frac{2\pi}{n})},\nonumber\\
\quad S_{ji}^{ij}(\theta)  &  =-F(\theta)\frac{\sin(\pi\lambda)\exp
(\mathrm{i}\kappa_{i,j}\lambda\theta)}{\sinh(\frac{n}{2}\lambda(\theta
-\mathrm{i}\frac{2\pi}{n})}, \label{S}%
\end{align}
where $i\neq j$ and $\kappa_{i,j}=i-j-\frac{n}{2}sgn(i-j).$ The function
$F(\theta)$ satisfies the unitarity and crossing symmetry conditions:%
\begin{equation}
F(\theta)F(-\theta)=1,
\qquad
F(\mathrm{i}\pi-\theta)F(\mathrm{i}\pi
+\theta)=\frac{\sinh(\frac{n}{2}\lambda(\theta+\mathrm{i}\pi))\sinh(\frac
{n}{2}\lambda(-\theta+\mathrm{i}\pi))}{\sinh(\frac{n}{2}\lambda(\theta
+\mathrm{i}\pi\frac{n-2}{n}))\sinh(\frac{n}{2}\lambda(-\theta+\mathrm{i}%
\pi\frac{n-2}{n}))} \label{u}%
\end{equation}
It is convenient to introduce parameter $p:$%
\[
p=\frac{1}{\lambda}%
\]
Then the solution to the equations (\ref{u}) can be written as:%
\begin{equation}
F_{n}(\theta)=\exp(\mathrm{i}\delta_{n,p}(\theta))=\exp\left(  \mathrm{i}%
{\displaystyle\int\limits_{-\infty}^{\infty}}\frac{d\omega}{\omega}\,
\frac{\sinh\left(  \frac{\pi(n-1)\omega}{n}\right)  \sinh\left(
\frac{\pi(p-1)\omega}{n}\right)  }
{\sinh(\pi\omega)\sinh\left(
\frac{\pi p\omega}{n}\right)  }\sin(\omega\theta)\right)  . \label{dn}%
\end{equation}
We have a one parameter scattering theory. To connect it with QFT
(\ref{acf}),(\ref{asl}) we should find a relation between parameter $\lambda$
(or $p$) and the coupling constant $b$. To solve this problem we need to
perform the non-perturbative Bethe Ansatz (BA) analysis.

\section{Bethe Ansatz analysis}

To find the relations between the parameters of the action and the scattering
parameters it is important to calculate the observables using both approaches.
Our QFT possesses $U(1)^{n}$ symmetry generated by the charges $Q_{j}=\int
\psi_{j}\psi_{j}^{\ast}dx.$ One can add to the Hamiltonian of our QFT the
terms $-A_{i}Q_{i}$ with different parameters $A_{i}$ (chemical potentials).
The simplest way to calculate the ground state energy using the BA approach is
to consider the configurations of the chemical potentials $A_{i}$ which lead
to the condensation of particles which have the simple scattering amplitudes,
namely, the pure phases. It is useful to consider the case when only one kind
of particles is condensed. This configuration corresponds to $A_{1}=A$ and all
other $A_{i}=0.$ In the case for $A>>M$ we can neglect in the action
(\ref{afb}) all terms containing the mass parameter $M_{0}$ and there only
particle $\psi_{1\text{ }}$will condense. It means that for calculation of
density of ground state energy (GSE) we can put all $\psi_{i}$ with $i>1$
equal to zero. The action in this case coincides with the action of massless
Thirring model:
\[
\mathcal{A}_{MTM}=\frac{1}{4\pi}\int d^{2}x\left[  \mathrm{i}\overline{\psi
}_{1}\gamma_{\mu}\partial_{\mu}\psi_{1}-\frac{(n-1)b^{2}}{n(1+b^{2}%
)}(\overline{\psi}_{1}\gamma_{\mu}\psi_{1})^{2}\right]  .
\]
The GSE of massless Thirring model with coupling constant $g=-\frac
{(n-1)b^{2}}{n(1+b^{2})}$ is well known and is given by%
\begin{equation}
\mathcal{E}_{MTM}\mathcal{(}A)=-\frac{1}{2\pi}(1+g)^{-1}=-\frac{n(1+b^{2}%
)}{2\pi(n+b^{2})}. \label{ass1}%
\end{equation}
We calculate now the same value from the BA approach. For $A>M$ one has the
sea of particles $\psi_{1}(\theta)$ which fill all possible states inside some
\textquotedblleft Fermi interval\textquotedblright\ $-B<\theta<B.$ The
distribution of the particles $\epsilon(\theta)$\ within this interval is
determined by their scattering amplitude $S_{11}^{11}(\theta)=-F_{n}(\theta).$
The specific ground state energy has the form
\begin{equation}
\mathcal{E(}A)\mathcal{-E(}0\mathcal{)=-}\frac{M}{2\pi}%
{\displaystyle\int\limits_{-B}^{B}}
\cosh(\theta)\epsilon(\theta)d\theta\label{en}%
\end{equation}
where the non-negative function $\epsilon(\theta)$ satisfies the BA equation,
inside the interval $-B<\theta<B,$
\begin{equation}%
{\displaystyle\int\limits_{-B}^{B}}
\widetilde{K}(\theta-\theta^{\prime})\epsilon(\theta^{\prime})d\theta^{\prime
}=A-M\cosh\theta\label{BA}%
\end{equation}
and the parameter $B$ is determined by the boundary condition $\epsilon(\pm
B)=0.$

The kernel $\widetilde{K}(\theta)$ in this equation is related to $\psi
_{1}\psi_{1}$ scattering phase $\delta_{n,p}(\theta)$ (\ref{dn})%
\begin{equation}
\widetilde{K}(\theta)=\delta(\theta)-\frac{1}{2\pi\mathrm{i}}\frac{d}{d\theta
}\log(F_{n}(\theta))=\delta(\theta)-\frac{1}{2\pi}\frac{d}{d\theta}%
\delta_{n,p}(\theta). \label{k1}%
\end{equation}
The Fourier transform $K(\omega)$\ of \ this kernel has the form%
\begin{equation}
K(\omega)=\frac{\sinh\left(  \frac{\pi\omega}{n}\right)  \sinh\left(
\frac{\pi(p+n-1)\omega}{n}\right)  }{\sinh(\pi\omega)\sinh\left(  \frac{\pi
p\omega}{n}\right)  }. \label{k2}%
\end{equation}
The main term of asymptotics of the function $\mathcal{E(}A\rightarrow\infty)$
can be expressed explicitly through the kernel $K(\omega)$ by the relation
\cite{FOZ}
\begin{equation}
\mathcal{E(}A\rightarrow\infty)=-\frac{A^{2}}{2\pi K(0)}=-\frac{npA^{2}}%
{2\pi(p+n-1)}. \label{ass2}%
\end{equation}
Comparing the Eqs (\ref{ass1}) and (\ref{ass2}), we find that
\begin{equation}
p=\frac{1}{\lambda}=(1+b^{2}) \label{pb}%
\end{equation}
in agreement with perturbation theory. We see that at $b\rightarrow
\infty,\lambda\rightarrow0$ and $S$-matrix (\ref{S}) tends to the scattering
matrix of $CP(n-1)$ model with fermion \cite{Abd}. So it is natural to call
our QFT as the dual theory to deformed $CP(n-1)$ model with fermion.

The term $\mathcal{E(}0\mathcal{)}$ in Eq. (\ref{en}) is the bulk vacuum
energy of QFT (\ref{afb}). It can be expressed trough the kernel $K(\omega)$
by the relation
\begin{equation}
\mathcal{E(}0\mathcal{)=-}\frac{M^{2}}{8}\left(  K(\omega)\cosh\left(
\frac{\pi\omega}{2}\right)  |_{\omega=\mathrm{i}}\right)  ^{-1}=\frac
{M^{2}\sin\left(  \frac{\pi p}{n}\right)  }{4\sin\left(  \frac{\pi}{n}\right)
\sin\left(  \frac{\pi(p-1)}{n}\right)  }. \label{e0}%
\end{equation}
For small $b$ the bulk vacuum energy will be%
\begin{equation}
\mathcal{E(}0\mathcal{)=}\frac{n M^{2}}{4\pi b^{2}}+\frac{M^{2}}{4}\cot\left(
\frac{\pi}{n}\right)  +O(b^{2}), \label{e0b}%
\end{equation}
the first term where is the contribution of Toda term in Eq.(\ref{acf}). The
second term comes from the renormalization of $M=M_{0}(b^{2})$ and from the
contribution of the vacuum energies of $n$ free fermions and $(n-1)$ bosonic
Toda particles with masses $M_{k}=2M\sin(\pi k/n),(k=1,..,[n/2]).$ The second
contribution can be easily calculated and is given by%
\begin{align}
\delta\mathcal{E(}0\mathcal{)}  &  \mathcal{=}-\frac{M^{2}}{4\pi}\left(
{\displaystyle\sum\limits_{k=1}^{[n/2]}}
4\sin(\pi k/n)^{2}\log\left(  \frac{M_{k}^{2}}{M^{2}}\right)  -(1+(-1)^{n}%
)\log\left(  \frac{M_{n/2}^{2}}{M^{2}}\right)  \right) \nonumber\\
&  =-\frac{M^{2}}{4\pi}\left(
{\displaystyle\sum\limits_{k=1}^{[n/2]}}
4\sin(\pi k/n)^{2}\log\left(  4\sin(\pi k/n)^{2}\right)  -(1+(-1)^{n}%
)\log(4)\right)  . \label{sum}%
\end{align}
The exact relation between the physical mass $M$ and parameter $M_{0}$ can be
calculated by the BA methods \cite{F2},\cite{Al} and has the form:%
\begin{equation}
M=M_{0}\frac{\Gamma\left(  \frac{p}{n}\right)  }{\Gamma\left(  \frac{1}%
{n}\right)  \Gamma\left(  \frac{p+n-1}{n}\right)  }=M_{0}\left(  1+\frac
{(\psi\left(  \frac{1}{n}\right)  -\psi(1))}{n}b^{2}+O(b^{4})\right)
\label{mm0}%
\end{equation}
It is easy to derive from Eqs.(\ref{e0b})-(\ref{mm0}) the agreement
with the perturbation theory and the sum rule:%
\begin{align*}
&  \frac{1}{\pi}\left(
{\displaystyle\sum\limits_{k=1}^{[n/2]}}
\sin(\pi k/n)^{2}\log(4\sin(\pi k/n)^{2})-\frac{1}{4}(1+(-1)^{n}%
)\log(4)\right) \\
&  =-\left(  \frac{\cot\left(  \frac{\pi}{n}\right)  }{4}+\frac{2(\psi\left(
\frac{1}{n}\right)  -\psi(1))}{n}\right)
\end{align*}
here $\psi(z)=\Gamma^{\prime}(z)/\Gamma(z).$ We note that QFT (\ref{acf})
provides us by example of theory which has no UV divergences in perturbation
theory in $b.$ This nice property is general for QFTs possessing the
\textquotedblleft nice\textquotedblright\ duality with asymptotically free SMs.

For $b\rightarrow\infty$ we, represent $\mathcal{E(}0\mathcal{)=}\frac
{M^{2}\sin\left(  \frac{\pi p}{n}\right)  }{4\sin\left(  \frac{\pi}{n}\right)
\sin\left(  \frac{\pi(p-1)}{n}\right)  }=\frac{M^{2}}{4}(\cot(\pi/n)+\cot(\pi
b^{2}/n)).$ The second term oscillates around zero with period $b^{2}/n.$ It
is reasonable to define the bulk ground state energy as the meaning value over
some interval much bigger then this period. With this definition we derive
that bulk ground state energy of $CP(n-1)$ model with fermion is given by:
$\mathcal{E(}0\mathcal{)=}\frac{M^{2}}{4}\cot(\pi/n),$ what agrees with large
$n$ expansion.

The integral BA equation can be studied using the generalized Winner-Hopf
method \cite{G}. We can go further in the UV ($\frac{A}{M}>>1$) analysis of
the function $\mathcal{E(}A\mathcal{)}$. In particular, the UV corrections to
Eq.(\ref{ass2}) have the form $\left(  \frac{M}{A}\right)  ^{2\nu_{j}},$ where
the exponents $\nu_{j}$ are defined (with the factor $1/$\textrm{i}) by the
zeroes of the kernel (\ref{k2}) in the upper half-plane. For integer $n$ the
zeroes at $\omega_{k}=$\textrm{i}$nk$ are cancelled by the denominator of the
kernel (this phenomenon is related with the cancellation of instanton
contribution by the fermion in Eq (\ref{ad})). The zeroes at $\omega_{j}%
=$\textrm{i}$\frac{nj}{(p+n-1)},(\nu_{j}=\frac{nj}{(p+n-1)})$ we call the
perturbative ones. For small $b,$ the corresponding exponents $2\nu_{j}%
=\frac{2nj}{(b^{2}+n)}$ appear after summation of logarithms in the
perturbative series. As a result the UV expansion has the form%
\[
\mathcal{E(}A\mathcal{)=}-\frac{npA^{2}}{2\pi(p+n-1)}\left(  1+\sum
_{k}^{\infty}d_{k}\left(  \frac{M}{A}\right)  ^{\frac{2nk}{(p+n-1)}}\right)
\]
All coefficients $d_{k}$ can be expressed trough the residues of the function
$\rho(\omega)=N(\omega)/N(-\omega),$ where $N(\omega)$\ is the function
analytical in the lower half plane which factorizes the kernel $K(\omega
)=(N(\omega)N(-\omega))^{-1}.$ In particular, the first coefficient%
\[
d_{1}=-\frac{2(p+n-1)\Gamma(\frac{-1}{p+n-1})\Gamma(\frac{2p+3n-2}%
{p+n-1})\Gamma(\frac{-p}{p+n-1})}{(p+n-2)\Gamma(\frac{p+n}{p+n-1})\Gamma
(\frac{2p+n-2}{p+n-1})\Gamma(\frac{p}{p+n-1})}\left(  \frac{\Gamma
(\frac{p+2n-1}{n})\Gamma(\frac{n+1}{n})}{2\Gamma(\frac{p+n}{n})}\right)
^{^{\frac{2n}{p+n-1}}}%
\]

The function $\mathcal{E(}A\mathcal{)}$ simplifies drastically in the scaling
limit $b^{2}=p-1>>1,\log(\frac{A}{M})>>1$ with $\log(\frac{A}{M})/p$ fixed.
This approximation is useful for the analysis of the UV behavior in dual
(sigma-model) representation of our QFT. The function $\mathcal{E(}%
A\mathcal{)}$ in this approximation (up to $\frac{1}{p^{2}}$) is%
\begin{equation}
\mathcal{E(}A\mathcal{)=-}\frac{nA^{2}}{2\pi}\left(  1-\frac{(n-1)}%
{p+n-1}\left(  \frac{1+q^{2}}{1-q^{2}}+\frac{2(n-1)q^{2}\log(\frac{p(1-q^{2}%
)}{ne^{1/2}})}{p(1-q^{2})^{2}}\right)  \right)  \label{ea}%
\end{equation}
here $q=\left(  \frac{M\Gamma(\frac{n+1}{n})}{2A}\right)  ^{\frac{n}{p+n-1}},$
and the omitted terms are $O(1/p^{3}).$ In the limit $b^{2}=p-1\rightarrow
\infty,$ we derive the known result \cite{BASSO} for $CP(n-1)$ with fermion%
\begin{equation}
\mathcal{E(}A\mathcal{)=-}\frac{nA^{2}}{2\pi}\left(  1-\frac{n-1}{n\log\left(
A/M\right)  }-\frac{(n-1)^{2}\log\left(  \log(A/M)\right)  +n(n-1)D_{n}}%
{n^{2}\log\left(  A/M\right)  ^{2}}+..\right)  \label{ea1}%
\end{equation}
where $D_{n}=\log\left(  \Gamma(\frac{n+1}{n})\right)  -\frac{\log(2)}%
{n}-\frac{(n-3)}{2n}.$

We note that UV behavior of function (\ref{ea}),(\ref{ea1}) is not
characteristic for asymptotically free sigma models. In these models GSE
$\mathcal{E(}A\mathcal{)}/A^{2}$ usually tends to infinity in the scaling
limit. A different behavior, depending on the chose of parameters $A_{i},$
appears in the models which have several $U(1)$ symmetries (see \cite{F}).
Here we show that similar phenomenon takes place in our QFT. The GSE
(\ref{ea1}) is related with $U(1)$ associated with fermion $\chi$ in the
action (\ref{ad}) or with field $\phi_{0}$ in the action (\ref{asl}).

As it was noted before the BA equations for the GSE in the fields $A_{i}$ is
easy to study when the scattering amplitudes of condensed particles are pure
phases. The corresponding situation is realized when we take $A_{1}=A,$ and
$A_{2}=-A.$ In this case the scattering amplitudes are given by%
\[
S_{11}^{11}(\theta)=S_{\overline{2}\overline{2}}^{\overline{2}\overline{2}%
}(\theta)=\exp(\mathrm{i}\delta_{n,p}(\theta)),\quad S_{1\overline{2}%
}^{1\overline{2}}(\theta)=S_{\overline{2}1}^{\overline{2}1}(\theta
)=\exp(\mathrm{i}\delta_{n,p}^{(1)}(\theta)).
\]
The function $\delta_{n,p}(\theta)$ is given by Eq.(\ref{dn}) and%
\[
\delta_{n,p}^{(1)}(\theta)=%
{\displaystyle\int\limits_{-\infty}^{\infty}}
\frac{d\omega\sinh\left(  \frac{\pi\omega}{n}\right)  \sinh\left(  \frac
{\pi(p-1)\omega}{n}\right)  }{\omega\sinh(\pi\omega)\sinh\left(  \frac{\pi
p\omega}{n}\right)  }\sin(\omega\theta).
\]
With these data we can write the system of BA integral equations for the
calculation of function $\mathcal{E}_{-}\mathcal{(}A\mathcal{)}$%
\[
\mathcal{E}_{-}\mathcal{(}A)\mathcal{-E(}0\mathcal{)=-}\frac{M}{2\pi}\left(
{\displaystyle\int\limits_{-B}^{B}}
\cosh(\theta)\epsilon_{1}(\theta)d\theta+%
{\displaystyle\int\limits_{-B}^{B}}
\cosh(\theta)\epsilon_{2}(\theta)d\theta\right)
\]
where the non-negative functions $\epsilon_{1}(\theta),\epsilon_{2}(\theta)$
satisfy inside the interval $-B<\theta<B,$ the BA equations%
\begin{align}%
{\displaystyle\int\limits_{-B}^{B}}
\widetilde{K}_{1}(\theta-\theta^{\prime})\epsilon_{1}(\theta^{\prime}%
)d\theta^{\prime}+%
{\displaystyle\int\limits_{-B}^{B}}
\widetilde{K}_{2}(\theta-\theta^{\prime})\epsilon_{2}(\theta^{\prime}%
)d\theta^{\prime}  &  =A-M\cosh\theta;\nonumber\\%
{\displaystyle\int\limits_{-B}^{B}}
\widetilde{K}_{1}(\theta-\theta^{\prime})\epsilon_{2}(\theta^{\prime}%
)d\theta^{\prime}+%
{\displaystyle\int\limits_{-B}^{B}}
\widetilde{K}_{2}(\theta-\theta^{\prime})\epsilon_{1}(\theta^{\prime}%
)d\theta^{\prime}  &  =A-M\cosh\theta\label{sy}%
\end{align}
and the parameter $B$ is determined by the boundary condition $\epsilon
_{1,2}(\pm B)=0.$

The kernels $\widetilde{K}_{1,2}(\theta)$ in this equation are%
\begin{equation}
\widetilde{K}_{1}(\theta)=\widetilde{K}(\theta)=\delta(\theta)-\frac{1}{2\pi
}\frac{d}{d\theta}\delta_{n,p}(\theta),\quad\widetilde{K}_{2}(\theta
)=-\frac{1}{2\pi}\frac{d}{d\theta}\delta_{n,p}^{(1)}(\theta).\quad\label{ks}%
\end{equation}
The Fourier transforms $K_{1,2}(\omega)$\ of these kernels have the form
\[
K_{1}(\omega)=K(\omega),\quad K_{2}\left(  \omega\right)  =-\frac{\sinh\left(
\frac{\pi\omega}{n}\right)  \sinh\left(  \frac{\pi(p-1)\omega}{n}\right)
}{\sinh(\pi\omega)\sinh\left(  \frac{\pi p\omega}{n}\right)  }%
\]
The eqs.(\ref{sy}) are symmetric with respect to $\epsilon_{1}(\theta
)\leftrightarrow\epsilon_{2}(\theta).$ It means that we can reduce them to one
equation with the kernel%
\[
K_{0}(\omega)=K(\omega)+K_{2}\left(  \omega\right)  =\frac{\sinh\left(
\frac{\pi\omega}{n}\right)  \cosh\left(  \frac{\pi(p+n/2-1)\omega}{n}\right)
}{\cosh(\frac{\pi\omega}{2})\sinh\left(  \frac{\pi p\omega}{n}\right)  }.
\]
This kernel $K_{0}(\omega)$ coincides exactly with the kernel which appears in
the BA equations for the deformed $O(n+2)$ sigma-model. It means that function
$\mathcal{E}_{-}\mathcal{(}A)\mathcal{-E(}0\mathcal{)}$ is equal twice the
function $\mathcal{E}_{O(n+2)}\mathcal{(}A)\mathcal{-E(}0\mathcal{)}$ for
deformed $O(n+2)$ sigma-model and possesses all characteristic properties for
asymptotically free sigma models. This function was studied in \cite{FL} and
in the scaling limit together with the first correction $O(1/p\log p)$ is%
\begin{equation}
\mathcal{E}_{-}\mathcal{(}A\mathcal{)}=2\mathcal{E}_{O(n+2)}\mathcal{(}%
A)\mathcal{=-}\frac{pA^{2}}{\pi}\frac{1-q}{1+q}\left(  1+\frac{2}{p}%
\frac{q\log\left(  \frac{(1-q^{2})p}{n}\right)  }{1-q^{2}}+..\right)  ,
\label{scf}%
\end{equation}
where $q=\left(  \frac{M}{A}\Gamma(\frac{n+1}{n})\right)  ^{\frac{n}{p}}.$ The
first term $\mathcal{-}\frac{pA^{2}}{\pi}\frac{1-q}{1+q}=\mathcal{-}%
\frac{pA^{2}}{\pi}\tanh\left(  \frac{n}{2p}\log\left(  \frac{M}{A}\right)
\right)  $ corresponds to the one loop sigma-model metric, whereas the next
term appears in the two-loop approximation.

\section{Sigma models, Ricci flow, integrability and observables}

In this section we consider the relation between integrable QFTs, scattering
theory data, classically integrable SMs and Ricci flow equations. Integrable
QFTs possess factorized scattering theory and their $S$-matrices satisfy
Yang-Baxter (YB) equations. When these QFTs have the dual SM representation,
the corresponding SMs should be also integrable at least classically. These
SMs are called YB SMs \cite{CL},\cite{DD}. It is very natural that these
objects are associated with names of Yang and Baxter. The scattering matrices
in the QFTs, which have the dual SM representation are the deformations of
$S$-matrices of SMs on the symmetric spaces (sometimes with additional degrees
of freedom which are necessary for quantum integrability). The YB SMs are also
the models on the deformed symmetric spaces. One can expect a correspondence
between the parameters of deformation in both theories. To find it, we compare
the observables calculated from scattering theory data and from SMs.

To realize this problem, it is very important to find a relation between YB
SMs and solutions of Ricci flow equations. In many cases this relation is
rather non-trivial and includes the diffeomorphisms of metrics depending on
the RG time (logarithm of the scale). In this section we discuss how sometimes
this problem can be avoided. We note that in general case the $S$-matrices,
corresponding to deformed symmetric spaces, do not have the values of the
deformation parameters where they are equal to identity matrice. In these
cases the representation of QFT in terms of fields does not appear to be very
useful, because there is no weak coupling region. However, even in these cases
such representation can be used for construction of non-local integrals of
motion, which help to reconstruct the $S$-matrix.

To start with, we consider the case where we have the exact correspondence
between QFT, factorized scattering theory (\ref{SSp}), and SM \cite{F}. All
these objects depend on two parameters $p_{1},p_{2}$ and describe a
two-parameter deformation of $SU\left(  2\right)  $ principle chiral field. To
write the action of this SM, it is useful to introduce the coordinates for
$g\subset SU\left(  2\right)  $%
\[
g=\left(
\begin{array}
[c]{cc}%
e^{\mathrm{i}\beta_{1}+\mathrm{i}\beta_{2}}\cos\theta & e^{\mathrm{i}\beta
_{1}-\mathrm{i}\beta_{2}}\sin\theta\\
e^{-\mathrm{i}\beta_{1}+\mathrm{i}\beta_{2}}\sin\theta & e^{-\mathrm{i}%
\beta_{1}-\mathrm{i}\beta_{2}}\cos\theta
\end{array}
\right)  .
\]
Then action is%
\begin{equation}
\mathcal{A}\left(  SU\left(  2\right)  \right)  =\frac{\sigma}{4\pi}\int
d^{2}x\left(  \frac{u(t)tr\left(  \partial_{\mu}g\partial_{\mu}g^{-1}\right)
-2l(t)(L_{\mu}^{3})^{2}-2r(t)(R_{\mu}^{3})^{2}}{\left(  a(t)^{2}%
-b(t)^{2}(tr\left(  g\tau_{3}g^{-1}\tau_{3}\right)  \right)  }\right)  ,
\label{tpfam}%
\end{equation}
where $L_{\mu}^{a}=\frac{1}{2}tr(\left(  \partial_{\mu}gg^{-1}\tau_{a}\right)
),$ $R_{\mu}^{a}=\frac{1}{2}tr(\left(  \partial_{\mu}g^{-1}g\tau_{a}\right)
),$ $\sigma=\frac{4p_{1}p_{2}}{(p_{1}+p_{2})}$,%
\begin{equation}
\begin{split}
u(t)  &  =\tau,
\qquad
l(t)+r(t)=\tau\left(  1+\sqrt{(1-\tau^{2})(1-\mathrm{k}%
^{2}\tau^{2})}\right)  ,\\
l(t)-r(t)  &  =\mathrm{k}\tau^{2},\qquad
a(t)+b(t)=\sqrt{1-\mathrm{k}^{2}\tau^{2}},
\qquad
a(t)-b(t)=\sqrt{1-\tau^{2}},
\end{split}
\end{equation}
and parameter $.$%
\begin{equation}
\mathrm{k}=\frac{p_{1}-p_{2}}{p_{1}+p_{2}}%
\end{equation}
The function $\tau(t)$ which was denoted in \cite{F} as $\tanh\xi,$ is defined
by the relation%
\begin{equation}
\left(  \frac{1-\tau}{1+\tau}\right)  \left(  \frac{1-\mathrm{k}\tau
}{1+\mathrm{k}\tau}\right)  ^{\mathrm{k}}=q^{2}(  t)  ,
\qquad
q(t)  =\exp\left(  \frac{2(t-t_{0})}{(p_{1}+p_{2})}\right)  \label{qtau}%
\end{equation}
or $\xi-\frac{\mathrm{k}}{2}\log\left(  \frac{1-\mathrm{k}\xi}{1+\mathrm{k}%
\xi}\right)  =\frac{2(t_{0}-t)}{(p_{1}+p_{2})}.$

The metric (\ref{tpfam}) in the coordinates $\chi_{1}=\frac{\beta_{1}%
+\beta_{2}}{\sqrt{2}},\chi_{2}=\frac{\beta_{1}-\beta_{2}}{\sqrt{2}}$ has the
form%
\begin{align}
ds^{2}  &  =\sigma\frac{u(d\theta)^{2}+\left(  u+r+l\cos2\theta\right)
(d\chi_{1})^{2}+\left(  u+l+r\cos2\theta\right)  (d\chi_{2})^{2}}{a^{2}%
-b^{2}\cos^{2}2\theta}\nonumber\\
&  +\sigma\frac{2\left(  u+l+r\right)  \cos2\theta d\chi_{1}d\chi_{2}}%
{a^{2}-b^{2}\cos^{2}2\theta}. \label{mt1}%
\end{align}
With this action, we can calculate the observables \cite{F}, The metric
(\ref{tpfam}) possesses two $U\left(  1\right)  $ symmetries. \ We introduce
the fields $A_{j},$ i.e. $\partial_{0}\chi_{j}\rightarrow(\partial_{0}%
+$\textrm{i}$A_{j})\chi_{j}.$ To compare the GSE, which is defined as the
minimum of the action of SM, with that derived in the previous section from
BA, we consider the same configurations of fields (chemical potentials) $A_{j}:$ $A_{1}=A,\ A_{2}=0$
and $A_{1}=-A_{2}=A.$ The field $A$ introduces the scale in our QFT. We put
\[
t_{0}-t=\log\frac{A}{M}.
\]
For the first configuration the minimum of the action corresponds to
$\theta=0$ and we have%
\begin{equation}
\mathcal{E=}-\frac{\sigma A^{2}(u+r+l)}{4\pi(a^{2}-b^{2})}=-\frac{2\sigma
A^{2}\tau(A)}{4\pi}.
\end{equation}
For the second the minimum of the action corresponds to $\theta=\frac{\pi}{4}$
and%
\begin{equation}
\mathcal{E}_{-}\mathcal{=}-\frac{2\sigma A^{2}(u+r)}{4\pi a^{2}}=-\frac{\sigma
A^{2}\tau(A)}{4\pi}\left(  1+\sqrt{(1-\tau^{2})(1-\mathrm{k}^{2}\tau^{2}%
)}-\mathrm{k}\tau^{2}\right)  ^{-1}. \label{e-}%
\end{equation}

It is useful to compare the metric (\ref{tpfam}) with integrable metric
\cite{CL}, which also depends on two parameters $\alpha$ and $\zeta$ and can
be written for arbitrary Lie group $G.$ This action can be written using the
operator $R$ acting on the generators $T_{a}$ of Lie algebra $G$ as follows:
it multiplies \ the generators corresponding to positive roots by
\textrm{i}$,$ to negative roots by $-$\textrm{i} and annihilates Cartan
subalgebra. Let $\mathcal{R}_{a}^{b}$ is defined as $R\left(  T_{a}\right)
=\mathcal{R}_{a}^{b}T_{b}$ and $A(g)$ is the matrix of \ group $G$ in the
adjoin representation. We denote $J_{a,\mu}=tr(g^{-1}\partial_{\mu}gT_{a})$
\begin{equation}
\mathcal{A}_{G}^{\prime}=\frac{\delta^{\mu,\nu}+\varepsilon^{\mu,\nu}}{4\pi
}\Sigma(G)\tau\int d^{2}x\left(  J_{a,\mu}\left(  I+\mathrm{i}\alpha
\mathcal{R+}\mathrm{i}\zeta A(g)\mathcal{R}A(g)^{-1}\right)  _{a,b}%
^{-1}J_{a,\nu}\right)  . \label{clim}%
\end{equation}
For the case of $G=SU(2)$, $\mathcal{R=}\left(
\begin{array}
[c]{ccc}%
{\small 0} & {\small 1} & {\small 0}\\
{\small -1} & {\small 0} & {\small 0}\\
{\small 0} & {\small 0} & {\small 0}%
\end{array}
\right)  $, $\Sigma(G)=\sigma\tau$,
\[
A(g)\mathcal{R}\ A(g)^{-1}=\left(
\begin{array}
[c]{ccc}%
0 & \cos2\theta & -\sin2\beta_{1}\sin2\theta\\
-\cos2\theta & 0 & \cos2\beta_{1}\sin2\theta\\
\sin2\beta_{1}\sin2\theta & -\cos2\beta_{1}\sin2\theta & 0
\end{array}
\right)
\]
and the metric $ds^{\prime2}$ in the coordinates $\theta,\chi_{1},\chi_{2}$ is%
\begin{align}
&  \sigma\tau\left(  \frac{d\theta^{2}+2\cos^{2}\theta(1-\left(  \alpha
+\zeta\right)  ^{2}\cos^{2}\theta)d\chi_{1}^{2}+2\sin^{2}\theta(1-\left(
\alpha-\zeta\right)  ^{2}\sin^{2}\theta)d\chi_{2}^{2}}{1-\alpha^{2}-\zeta
^{2}-2\alpha\zeta\cos2\theta}\right) \nonumber\\
&  +\sigma\tau\left(  \frac{(\zeta^{2}-\alpha^{2})\sin^{2}2\theta d\chi
_{1}d\chi_{2}}{1-\alpha^{2}-\zeta^{2}-2\alpha\zeta\cos2\theta}\right)  .
\label{mt2}%
\end{align}

The metrics (\ref{mt1}) and (\ref{mt2}) are rather different. However, if we
take%
\begin{equation}
\alpha=\frac{1}{2}(1+\mathrm{k})\tau,\ \zeta=\frac{1}{2}(1-\mathrm{k}%
)\tau\label{azrel}%
\end{equation}
the observables calculated \ with these metrics coincide. For the first
configuration the minimum of action (\ref{mt2}) corresponds to $\theta=0$ and
is%
\[
\mathcal{E}\left(  A\right)  \mathcal{=}-\frac{2\sigma A^{2}\tau(A)}{4\pi}.
\]
For other configuration, the position of the minimum is less trivial%
\[
\cos^{2}\theta_{-}=\frac{1-(\zeta^{2}-\alpha^{2})-\sqrt{1-2\zeta^{2}%
-2\alpha^{2}+(\zeta^{2}-\alpha^{2})^{2}}}{\zeta\alpha}..
\]
For these values of $\theta_{-}$ and relations (\ref{azrel}), the energies
$\mathcal{E}_{-}$ coincide with (\ref{e-}).

The metric (\ref{mt2}),(\ref{azrel}) does not satisfy the Ricci flow equations
with coordinates independent on RG time $t$ but there is a transformation of
coordinates depending on $t$ which leads to the solution of these equations.
This transformation from the metric (\ref{mt2}) to (\ref{mt1}) was constructed
in \cite{TS}.

For arbitrary group $G$ we can calculate the observable with the action
(\ref{clim}) and $\alpha,\zeta$ defined by (\ref{azrel}) where in relation
(\ref{qtau}) defining $\tau(t)$ we should take $q(t)=q_{G}(t)=\exp\left(
\frac{h^{\vee}(G)(t-t_{0})}{(p_{1}+p_{2})}\right)  ,$ where $h^{\vee}(G)$ is
the dual Coxeter number. The factor $\Sigma(G)$ is simply related with the
kernel $K_{G}\left(  \omega\right)  $ in the BA equations associated with the
fundamental representation $~$of $G.$ $\Sigma(G)=\frac{2}{K_{G}(0)}.$ For
$G=SU\left(  n\right)  $ this kernel is\footnote{The scattering matrix of basic particles, i.e. particles with minimal mass in
deformed principle chiral SM, $S(G)_{p_{1},p_{2}}\left(  \theta\right)  $ is
equal up to a simple CDD factor to the direct product of the well known
$G_{\mathrm{q}}$ symmetric $S$-matricies $S(G)_{\mathrm{q}}$ in the fundamental (vector)
representation of $G$: $S(G)_{p_{1},p_{2}}=S(G)_{\mathrm{q}_{1}}\otimes
S(G)_{\mathrm{q}_{2}}$ with $\mathrm{q}_{i}=\exp(\mathrm{i}\frac{2\pi}{p_{i}%
}).$}
\begin{equation}
K_{SU\left(  n\right)  }=\frac{\sin\left(  \frac{\pi\omega}{n}\right)
\sin\left(  \frac{\left(  n-1\right)  \pi\omega}{n}\right)  \sin\left(
\frac{\left(  p_{1}+p_{2}\right)  \omega}{n}\right)  }{\sin\left(  \pi
\omega\right)  \sin\left(  \frac{p_{1}\omega}{n}\right)  \sin\left(
\frac{p_{2}\omega}{n}\right)  },
\qquad
\Sigma(SU\left(  n\right)  )=\frac
{2np_{1}p_{2}}{\left(  n-1\right)  \left(  p_{1}+p_{2}\right)  } \label{ksun}%
\end{equation}
The GSE $\mathcal{E}\left(  A\right)  $ for the field $A$ coupled with $U(1)$
generated by one of generators $H_{i}$ from Cartan subalgebra for $SU\left(
n\right)  ,$ is%
\begin{equation}
\mathcal{E}\left(  A\right)  =-\frac{np_{1}p_{2}}{2\pi\left(  n-1\right)
\left(  p_{1}+p_{2}\right)  }A^{2}\tau(A). \label{etau}%
\end{equation}
Comparing the kernels (\ref{ksun}) and (\ref{k2}), we see that $p_{1}%
=n-1,p_{2}=p.$ In the scaling limit $p\gg1,$ we can solve the equation
(\ref{qtau}) by perturbation theory in $\frac{1}{p}.$ In the first order we
derive ($\left(  t_{0}-t\right)  =\log\frac{A}{M})$)%
\begin{equation}
\tau(A)=1-\frac{2(n-1)}{p}\frac{q^{2}}{1-q^{2}}+\cdots ,
\qquad
q(A)=q_{SU(n)  }(A)=\exp\left(  \frac{n\log\frac{A}{M}}{(p_{1}+p_{2})}\right)
\label{tA}%
\end{equation}
It means that at the first order in $\frac{1}{p},$ the GSE $\mathcal{E}\left(
A\right)  $ defined by (\ref{etau}) is%
\begin{equation}
\mathcal{E}\left(  A\right)  =-\frac{n}{2\pi}\left(  1-\frac{n-1}{p}\left(
\frac{1+q^{2}}{1-q^{2}}\right)  \right)  +O(1/p^{2})
\end{equation}
in exact agreement with eq (\ref{ea}) derived in the BA approach. 

We consider the function $\mathcal{E}_{-}\left(  A\right)  $ for $G=SU\left(  2\right)  $
(\ref{e-}) studied by the BA method in the previous section. Using relations
(\ref{e-},\ref{tA}) we obtain in the leading order in $\frac{1}{p}$%
\begin{equation}
\mathcal{E}_{-}\left(  A\right)  =-\frac{A^{2}}{\pi}p\left(  \frac{1-q}%
{1+q}\right)  +O\left(  1\right)  \label{eq-}%
\end{equation}
in exact agreement with (\ref{scf}) for $n=2$.

It is useful to consider the metric (\ref{mt1}), which is the exact solution
of Ricci equations in the limit $p_{1}\ll p_{2}=p.$ In our case $p_{1}=1.$ we
derive from (\ref{mt1},\ref{tA}) that%
\begin{equation}
ds^{2}=4p\left(  \frac{1-q}{1+q}\right)  \frac{d\theta^{2}+\sin^{2}%
2\theta(d\beta)^{2}}{\left(  1-\left(  \frac{1-q}{1+q}\right)  ^{2}\cos
^{2}2\theta\right)  }+O(1) \label{dsu2}%
\end{equation}
This metric is the sausage metric and describes the deformed $CP(1)$ model.
The energy $\mathcal{E}_{-}\left(  A\right)  $ describes the degrees of
freedom related with SM. The parameter $p_{1}\ll p_{2}=p$ does not appear in
this limit. This parameter is related with additional degrees of freedom which
are necessary for the quantum integrability of QFT.

The main term of the metric (\ref{dsu2}) (sausage metric) can be derived by
the reduction procedure for coset $\frac{SU\left(  2\right)  }{U(1)}.$ The
deformed $CP(n-1)$ SM can be considered as SM on the coset space
$\frac{SU\left(  n\right)  }{SU\left(  n-1\right)  U(1)}$. In general the
action of SM on the coset space \cite{DD} can be written as%
\begin{equation}
\mathcal{A}_{c}=\frac{\delta^{\mu\nu}+\varepsilon^{\mu\nu}}{4\pi
}{\small \Sigma\kappa}\int d^{2}x\left(  J_{a,\mu}\left(
I\mathcal{+\mathrm{i}\kappa}A(g_{c})\mathcal{R}A(g_{c})^{-1}P\right)
_{a,b}^{-1}J_{a,\nu}\right)  , \label{accos}%
\end{equation}
where $g_{c}$ is the coset element of the group from the numerator of coset,
$P$ is the projector operator which annihilates the currents from denominator
of coset and ${\small \Sigma}$ the numerical factor.

To simplify the expressions we consider here the case $n=3,$ i.e. the deformed
$CP\left(  2\right)  $ SM (the same consideration can be done for any
$n,$\ using eq(\ref{accos}) and results of appendix B). The matrix $g\subset
SU\left(  3\right)  $ can be taken as (see appendix B)%
\begin{align}
g  &  ={\small e}^{\mathrm{i}\lambda_{3}\eta_{1}+\mathrm{i}\lambda_{8}\eta
_{8}}\left(
\begin{array}
[c]{ccc}%
{\small \cos\theta}_{3} & {\small 0} & {\small e}^{-\mathrm{i}\beta_{3}%
}{\small \sin\theta}_{3}\\
{\small 0} & {\small 1} & {\small 0}\\
{\small -e^{\mathrm{i}\beta_{3}}\sin\theta}_{3} & {\small 0} & {\small \cos
\theta}_{3}%
\end{array}
\right)  \left(
\begin{array}
[c]{ccc}%
{\small \cos\theta}_{1} & {\small e}^{-\mathrm{i}\beta_{1}}{\small \sin\theta
}_{1} & {\small 0}\\
{\small -e^{\mathrm{i}\beta_{1}}\sin\theta}_{1} & {\small \cos\theta}_{1} &
{\small 0}\\
{\small 0} & {\small 0} & {\small 1}%
\end{array}
\right) \nonumber\\
& \times \left(
\begin{array}
[c]{ccc}%
{\small 1} & {\small 0} & 0\\
{\small 0} & {\small \cos\theta}_{2} & {\small e}^{-\mathrm{i}\beta_{2}%
}{\small \sin\theta}_{2}\\
{\small 0} & {\small -e^{\mathrm{i}\beta_{2}}\sin\theta}_{2} & {\small \cos
\theta}_{2}%
\end{array}
\right)  \label{gsu3}%
\end{align}
the coset element%
\begin{equation}
g_{c}=\left(
\begin{array}
[c]{ccc}%
{\small e}^{\mathrm{i}\beta_{1}}{\small \cos\theta}_{1} & {\small e}%
^{\mathrm{i}\beta_{2}}{\small \cos\theta}_{2}{\small \sin\theta}_{1} &
{\small \sin\theta}_{1}{\small \sin\theta}_{2}\\
{\small -\sin\theta}_{1} & {\small e}^{\mathrm{i}\beta_{2}-\mathrm{i}\beta
_{1}}{\small \cos\theta}_{2}{\small \cos\theta}_{1} & {\small e}%
^{-\mathrm{i}\beta_{1}}{\small \sin\theta}_{2}{\small \cos\theta}_{1}\\
{\small 0} & -{\small \sin\theta}_{2} & {\small e}^{-\mathrm{i}\beta_{2}%
}{\small \cos\theta}_{2}%
\end{array}
\right)  \label{gco}%
\end{equation}
In terms of these coordinates, we can calculate the metric $ds^{2}$ and the
field $B.$ We note that our theory for $n>2$ possesses only $\mathbf{PT}$
symmetry so that the $B$-field should appear. We introduce the notations%
\begin{align*}
\mathrm{s}(  {\small \theta}_{1},{\small \theta}_{2})   &
=1-2\cos^{2}{\small \theta}_{2}\sin^{2}{\small \theta}_{1},
\\
\mathrm{Z}(  {\theta}_{1},{\theta}_{2},\kappa)   &
=(1-\kappa^{2}\cos^{2}(2{\small \theta}_{2}))\left(  1-\kappa^{2}\mathrm{s}%
\left(  {\small \theta}_{1},{\small \theta}_{2}\right)  \right)  +\kappa
^{2}\cos^{4}{\small \theta}_{2}\sin^{2}(2{\small \theta}_{1})\sin^{2}{\small \theta}_{2}.
\end{align*}
Then we find that 
\begin{align}
ds^{2}=&  \frac{\Sigma\kappa}{\mathrm{Z}(  {\theta}_{1},{\theta}_{2},\kappa)}\left\{
\cos^{2}{\small \theta}_{2}\left(  1-\kappa^{2}\cos^{2}(2{\small \theta}_{2})\right)  
(d\theta_{1})^{2}+\frac{\kappa^{2}}{2}\sin(4\theta_{2})
\sin(2{\small \theta}_{1})d\theta_{1}d\theta_{2}
\right.
\label{cp2m}\\
&  +\left(  1-\kappa^{2}(\mathrm{s}\left(  {\small \theta}_{1},
{\small \theta}_{2}\right)  -\cos^{4}{\small \theta}_{2}\sin^{2}(2{\small \theta}_{1})
\sin^{2}{\small \theta}_{2})\right)  (d\theta_{2})^{2}
\nonumber\\
& + \cos^{2}{\small \theta}_{2}\cos^{2}{\small \theta}_{1}
\Bigl[  \left(  1-\cos^{2}{\small \theta}_{2}\cos^{2}{\small \theta}%
_{1}\right)  -\kappa^{2}\left(  \cos^{2}(2{\small \theta}_{2})
\sin^{2}{\small \theta}_{1}+\cos^{2}{\small \theta}_{1}\sin^{2}{\small \theta}%
_{2}\right)  \Bigr]  (d\beta_{1})^{2}\nonumber\\
& \left.
- \frac{1}{2}\sin^{2}(2{\small \theta}_{2})\cos^{2}{\small \theta}_{1}\left(1-\kappa^{2}\mathrm{s(}{\small \theta}_{1},{\small \theta}_{2})\right)
d\beta_{1}d\beta_{2}
+\frac{1}{4}\sin^{2}(2{\small \theta}_{2})\left(1-\kappa^{2}\mathrm{s}^{2}\mathrm{(}{\small \theta}_{1},{\small \theta}_{2})\right)  (d\beta_{2})^{2}
\right\}
\nonumber
\end{align}
and the field 
\begin{align}
B=&  \frac{\mathrm{i}\Sigma\kappa^2}{\mathrm{Z}(  {\theta}_{1},{\theta}_{2},\kappa)}\left\{
\sin(2\theta_{1})\cos^{2}{\small \theta}_{2}\left(  \left(  1-\kappa^{2}%
\cos^{2}(2{\small \theta}_{2})\right)  \mathrm{s}\left(  {\small \theta}_{1},{\small \theta}_{2}\right)  
-\frac{\sin^{2}(2{\small \theta}_{2})}{2}%
\cos^{2}{\small \theta}_{1}\right)  d\theta_{1}\wedge d\beta_{1}
\right.
\nonumber\\
&  +\frac{\sin(2\theta_{1})}{2}\cos^{2}{\small \theta}_{2}\sin^{2}(2{\small \theta}_{2})
d\theta_{1}\wedge d\beta_{2}-\left(  1-\kappa^{2}\mathrm{s(}%
{\small \theta}_{1},{\small \theta}_{2})\right)  \frac{\sin(4\theta_{2})}{2}%
\cos^{2}{\small \theta}_{2}d\theta_{2}\wedge d\beta_{1}\nonumber\\
&\left.  +\frac{1}{2}\sin(4\theta_{2})\left(  1-\kappa^{2}\mathrm{s}^{2}%
\mathrm{(}{\small \theta}_{1},{\small \theta}_{2})\right)  d\theta_{2}\wedge
d\beta_{2}\right\}. 
\label{Bcp2}%
\end{align}

The metric (\ref{cp2m}) and the field $B$ (\ref{Bcp2}) satisfy Ricci flow
equations for ${\small \Sigma=}\frac{1}{\nu}$ and $\kappa\left(  t\right)
=\tanh\left(  6\nu\left(  t_{0}-t\right)  \right)  .$ To show this, we use the
conjecture, which is to be proven in the next paper, namely that Ricci flow
equations are satisfied if they are satisfied for abelian $T$-dual SM. For
$T$-dual SM with respect parameters $\beta_{1},\beta_{2},$ the field $B$
disappears and the metric simplifies:%
\begin{equation}
\begin{split}
ds_{dual}^{2}  &  =\kappa(t)\left(  \cos\theta_{1}d\theta_{2}+\frac
{\mathrm{i\tan}\theta_{2}\left(  \cos2\theta_{1}-\cot^{2}\theta_{2}\right)
d\beta_{1}-\mathrm{i}\cot\theta_{2}d\beta_{2}}{\cos\theta_{1}}\right)  ^{2}\\
& + \kappa(t)(d\theta_{1}-2\mathrm{i\cot}2\theta_{1}d\beta_{2})^{2}%
+\frac{\left(  \left(  \frac{d\beta_{1}}{\cos\theta_{2}}+\cos\theta_{2}%
d\beta_{2}\right)  ^{2}+\frac{\sin^{2}\theta_{2}}{\sin^{2}\theta_{1}}%
d\beta_{2}^{2}\right)  }{\kappa(t)\sin^{2}\theta_{2}\cos^{2}\theta_{2}}%
\end{split}
\end{equation}
It is easy to check that Ricci flow equations%
\begin{equation}
\frac{d}{dt}g_{ij}=-R_{ij}+2\nabla_{i}\nabla_{j}\Phi
\end{equation}
are satisfied with dilaton field $\Phi=\log\left(  \cos\theta_{1}\sin
2\theta_{1}\sin2\theta_{2}\right)  +4\mathrm{i}\left(  \beta_{1}+2\beta_{2}\right)  $ and 
\begin{equation}
\kappa(t)=\frac{\tanh\left(  6\nu\left(  t_{0}-t\right)
\right)  }{\nu}
\end{equation}

To compare the observables calculated from the SM representation with that's
derived from scattering theory data by the BA method, we introduce the fields
related with two $U\left(  1\right)  $ symmetries of SM (\ref{cp2m}),(\ref{Bcp2}). 
Namely $\partial_{0}\beta_{i}\rightarrow(\partial_{0}+$\textrm{i}$A_{i})\beta_{i}$. 
The configuration considered by the BA method
corresponds $A_{1}=-A_{2}=A.$ In field $A,$ $\left(  t_{0}-t\right)
=\log\frac{A}{M}.$ It is easy to find that the minimum of the SM action
corresponds to $\theta_{1}=0,\theta_{2}=\frac{\pi}{4}.$ The minimum of the
action defines the ground state energy $\mathcal{E}_{-}\left(  A\right)  $ for
the deformed SM and is given by%
\begin{equation}
\mathcal{E}_{-}\left(  A\right)  =-\frac{A^{2}}{4\pi\nu}\tanh\left(  6\nu
\log\frac{A}{M}\right)  . \label{cp2e-}%
\end{equation}
For deformed $CP\left(  1\right)  $ SM we had from eq(\ref{eq-}) that%
\[
\mathcal{E}_{-}\left(  A\right)  =-\frac{A^{2}}{\pi}p\tanh\left(  \frac{1}%
{p}\log\frac{A}{M}\right)  =-\frac{A^{2}}{4\pi\nu}\tanh\left(  4\nu\log
\frac{A}{M}\right)  .
\]
Comparing these equations it is natural to assume that $\nu=\frac{1}{4p}$ and
for the deformed $CP\left(  n-1\right)  $ SMs one obtains%
\begin{equation}
\mathcal{E}_{-}\left(  A\right)  =-\frac{A^{2}}{\pi}p\tanh\left(  \frac{n}%
{2p}\log\frac{A}{M}\right)
\end{equation}
in exact agreement with the BA calculations (\ref{scf}). The parameter ${\small \Sigma
=}\frac{1}{\nu}=4p,$ $p=1+b^{2}$ (\ref{pb}). It provides a relation between
the parameters of QFTs (\ref{asl}),(\ref{acf}) and sigma models.

\section{Conformal limit of metric, semiclassical reflection amplitudes and UV
asymptotic of central charge}

In the previous section we conjectured a duality between the deformed
$CP(n-1)$ SM with fermion and QFT described by the action (\ref{asl}%
),(\ref{afb}). Here we consider a duality between the CFT with action
(\ref{ac1}) and conformal limit of the metric of deformed $CP(n-1)$ SM. The
free term in action $\mathcal{A}_{CFT}$ corresponds to fermion. \ For
simplification we restrict ourselves to the case $n=3.$ The case $n=2$,
corresponding to the duality between the Sine-Liouville CFT and Witten's $2d$
black hole SM, was considered in \cite{FS}.

To derive the conformal limit of the metric $ds^{2}$ and the $B$-field, we use
the method of contraction \cite{FL}. We denote as $u$ the parameter%
\begin{equation}
u=\frac{1}{p}(t_{0}-t)=\frac{1}{1+b^{2}}(t_{0}-t).
\label{defu}
\end{equation}
In UV limit $u\rightarrow\infty,\kappa(u)=\tanh\frac{3}{2}u\rightarrow1.$ We
introduce the parameter $\delta/2=1-\tanh u\rightarrow0$ and make the
rescaling of variables $\vartheta_{1}\rightarrow\sqrt{\delta}/2\vartheta
_{1},\vartheta_{2}\rightarrow\sqrt{\delta}/2\vartheta_{2}$ with $\delta
\rightarrow0.$ As a result, the conformal limit of the metric $ds^{2}$ can be
written as ($\frac{1}{\nu}=4p$)%
\begin{align}
ds_{C}^{2}  &  =p\frac{\left(  1+\theta_{2}^{2}\right)  (d\theta_{1}%
)^{2}+\left(  1+\theta_{1}^{2}\right)  (d\theta_{2})^{2}+\theta_{1}\theta
_{2}d\theta_{1}d\theta_{2}}{\left(  1+\theta_{1}^{2}+\theta_{2}^{2}+\frac
{3}{4}\theta_{1}^{2}\theta_{2}^{2}\right)  }
\label{ds2c}\\
& + p\frac{(\theta_{1}^{2}+\theta_{2}^{2}+\theta_{1}^{2}\theta_{2}^{2}%
)(d\beta_{1})^{2}+\theta_{2}^{2}\left(  1+\theta_{1}^{2}\right)  (d\beta
_{2})^{2}-(2+\theta_{1}^{2})\theta_{2}^{2}d\beta_{1}d\beta_{2}}{\left(
1+\theta_{1}^{2}+\theta_{2}^{2}+\frac{3}{4}\theta_{1}^{2}\theta_{2}%
^{2}\right)  }\nonumber
\end{align}
and the field $B$%
\begin{align}
B_{C}  &  =\mathrm{i}p\frac{\theta_{1}\left(  1+\frac{\theta_{2}^{2}}%
{2}\right)  d\theta_{1}\wedge d\beta_{1}-\theta_{2}\left(  1+\frac{\theta
_{1}^{2}}{2}\right)  d\theta_{2}\wedge d\beta_{1}}{\left(  1+\theta_{1}%
^{2}+\theta_{2}^{2}+\frac{3}{4}\theta_{1}^{2}\theta_{2}^{2}\right)
}\nonumber\\
& - \mathrm{i}p\frac{\frac{1}{2}\theta_{1}\theta_{2}^{2}d\theta_{1}\wedge
d\beta_{2}+\theta_{2}\left(  1+\theta_{1}^{2}\right)  d\theta_{2}\wedge
d\beta_{2}}{\left(  1+\theta_{1}^{2}+\theta_{2}^{2}+\frac{3}{4}\theta_{1}%
^{2}\theta_{2}^{2}\right)  } \label{bcft}%
\end{align}
The action of SM with the metric (\ref{ds2c}) and the $B$-field (\ref{bcft})
can be written in terms of two complex scalar fields $\omega_{1}\left(
x\right)  =\theta_{1}e^{\mathrm{i}\beta_{1}},\omega_{1}\left(  x\right)
=\theta_{2}e^{\mathrm{i}\beta_{2}}.$ It describes CFT corresponding coset
$\frac{SL(3)_{K}}{SL(2)_{K}U\left(  1\right)  }$\ with $K=3+b^{2}.$ This CFT
can be used for construction of string theory moving on such coset manifold.

To identify our conformal SM with CFT (\ref{ac1}), we show that the reflection
amplitudes derived in the minisuperspace approach in our conformal SM coincide
with the semiclassical limit of the reflection amplitudes (\ref{Rs}). Here we
consider our SM on a circle of length $R.$ We have a scale and $(t_{0}%
-t)=\log\left(  \frac{1}{MR}\right)  $. In the UV scaling regime $\log\left(
\frac{1}{MR}\right)  \rightarrow\infty,\frac{1}{p}\rightarrow0,$ with $u$
defined in (\ref{defu}), our approximation is exact up to $O\left(  \frac{1}{p}\log p\right)  .$
In this approximation we can use minisuperspace approach to calculate ground
state energy $E_{0}\left(  R\right)  $, effective central charge $E_{0}(R)=-\frac{\pi c_{\rm eff}( R)  }{R}$ 
and energies $E_{i}(R)$ of exited states. Let $c_{\rm eff}(R)  =c_{UV}%
-e_{0}(R),E_{i}(  R)  =\frac{\pi e_{i}}{R}.$ \ Then the
minisuperspace equation for the spectrum and eigenfunctions has the form
\cite{FL}%
\begin{equation}
\left(-\nabla^{2}+\frac{1}{4}\mathcal{R}+\frac{1}{12}\mathcal{H}^{2}\right)\Psi_{i}%
=\frac{e_{i}}{6}\Psi_{i} \label{HT}%
\end{equation}
where $\mathcal{R}$ is the curvature and 3-form $\mathcal{H}$ is defined in
terms of two form $B$ as $\mathcal{H=}dB.$

The operator in the l.h.s. of equation (\ref{HT}) can be written as
$p\mathrm{H,}$ where the operator $\mathrm{H}$\ depends only on $u.$ It means
that \textrm{e}$_{i}=\frac{e_{i}}{p}$ are functions only on $u.$ A solution of
this equation with the metric and the $B$-field for the deformed $CP(n-1)$ SM
is accessible only numerically. Even for $n=2$ where $B=0$, eq. (\ref{HT})
reduces to Lam\'{e} equation which can be solved analytically only in the limits of
small and large values of the scaling variable $u$. For $u<<1$ the metric and
$\mathcal{R}$ reduce to those of $CP(n-1)$ and the spectrum is trivial. The
corrections in $u$ can be derived by the standard perturbation theory. In
particular,%
\begin{equation}
\mathrm{e}_{0}(u)=p(c_{UV}-c_{\rm eff}(  R)  )=\frac{3n}{2u}(1+O(u^{4})).
\label{eoir}%
\end{equation}

For $u>>1$ the calculations are more involved. The starting point is the
conformal limit of the action of SM. Here we consider the case $n=3,$ where
this action is given by (\ref{ds2c}), (\ref{bcft}).

It is convenient to parametrize the spectrum of the eq (\ref{HT}) as
$\mathrm{e}_{0}=12P^{2}=12(P_{1}^{2}+P_{2}^{2}).$ The functions $\Psi_{i}(\det
G)^{1/4},$ which are integrated with trivial measure, will be denoted as
$\Psi_{P}.$

The Weyl group $\mathbf{w}_{3}$ of the Lie algebra of $sl(3)$ contains $6$
elements and acts to the parameters $\left(  P_{1},P_{2}\right)  $ as
rotations in this plane by $\frac{2\pi l}{3}$ and reflection $P\rightarrow-P$. 
We denote $P_{s}=\widehat{s}P$. For calculation of the effective central
charge, we can consider functions $\Psi_{P}$, independent of $\beta_{1},\beta_{2}$. To simplify the notations and equations, we consider here the
isotopic symmetric sector where $\Psi_{P}=\Psi_{P}(\theta_{1},\theta_{2})$. It is convenient to make the
substitution $\theta_{i}=\sqrt{Y_{i}}$ and $\Psi_{P}=F\left(  Y_{1},Y_{2}\right)$, 
the we obtain from eq(\ref{HT}) that
\begin{align}
0  &  =\left(  Y_{1}F_{Y_{1}}+Y_{2}F_{Y_{2}}+Y_{1}\left(  1+Y_{1}\right)
F_{Y_{1},Y_{1}}+Y_{2}\left(  1+Y_{2}\right)  F_{Y_{2},Y_{2}}+Y_{1}%
Y_{2}F_{Y_{1},Y_{2}}\right) \nonumber\\
&  +\frac{1}{8Y_{1}Y_{2}} \left(  2Y_{1}+2Y_{1}+4(P_{1}^{2}+P_{2}^{2}\right)
)F. \label{dife}%
\end{align}
This equation can be soled by the Mellin transform method.%
\[
F\left(  Y_{1},Y_{2}\right)  =\int_{C}Y_{1}^{\mathrm{i}Q_{1}}Y_{2}%
^{\mathrm{i}Q_{2}}\widetilde{F}(Q_{1,}Q_{2})dQ_{1}dQ_{2},
\]
where the contour $C$ includes the poles of $\widetilde{F}(Q_{1,}Q_{2})$\ on
the real axe $\widetilde{F}(Q_{1,}Q_{2}).$ The equation (\ref{dife}) can be
reduced to a functional equation to $\widetilde{F}(Q_{1,}Q_{2})$%
\begin{align*}
\left(  \frac{P^{2}}{2}-Q_{1}^{2}-Q_{2}^{2}+Q_{1}Q_{2}\right)  \widetilde
{F}(Q_{1,}Q_{2})  &  =\left(  \frac{1}{2}-\mathrm{i}Q_{1}\right)
^{2}\widetilde{F}(Q_{1}-\mathrm{i},Q_{2})\\
&  +\left(  \frac{1}{2}-\mathrm{i}Q_{2}\right)  ^{2}\widetilde{F}(Q_{2}%
,Q_{2}-\mathrm{i}).
\end{align*}
A solution to this equation can be found in a similar way as it was done in
\cite{OP} for $sl\left(  3\right)  $ Toda equation. Namely, let $h_{i}$ be the
weights of the fundamental representation. Then%
\[
\widetilde{F}(Q_{1,}Q_{2})=\frac{\Gamma^{2}(\frac{1}{2}-\mathrm{i}Q_{1}%
)\Gamma^{2}(\frac{1}{2}-\mathrm{i}Q_{2})%
{\displaystyle\prod\limits_{j=1}^{3}}
\Gamma(\mathrm{i}Q_{1}+\mathrm{i}h_{j}\cdot P)\Gamma(\mathrm{i}Q_{2}%
-\mathrm{i}h_{j}\cdot P)}{\Gamma\left(  \mathrm{i}Q_{1}+\mathrm{i}%
Q_{2}\right)  }%
\]
This function is Weyl invariant. The contributions to the asymptotics of the
function $F\left(  Y_{1},Y_{2}\right)  $ give the poles at $Q_{1}=-h_{j}\cdot
P,Q_{2}=h_{m}\cdot P$ with $m\neq j.$ If we take into account that the roots
of $sl(n)$ are $h_{j}-h_{m},m\neq j,$ and denote $Y_{1}=e^{e_{1}\cdot X}%
,Y_{2}=e^{e_{2}\cdot X},$ we find that in the Weyl chamber the function
$F\left(  Y_{1},Y_{2}\right)  $ has the asymptotics%
\begin{equation}
F\left(  e^{e_{1}\cdot X},e^{e_{2}\cdot X}\right)  =\sum_{w_{3}}A\left(
P_{s}\right)  e^{\mathrm{i}P_{s}\cdot X},
\end{equation}
where%
\begin{equation}
A\left(  P\right)  =\Gamma^{2}\left(\frac{1}{2}+\mathrm{i}h_{1}\cdot P\right)
\Gamma^{2}\left(\frac{1}{2}-\mathrm{i}h_{3}\cdot P\right)%
{\displaystyle\prod\limits_{\alpha>0}}
\Gamma(-e_{\alpha}\cdot P).
\end{equation}
We see that the function $A\left(  P\right)  $ coincides with scaling, or
semiclassical limit ( i.e. $b>>1,$ $\mathbf{a}b=$\textrm{i}$P$ is fixed) of
quantum amplitude $\mathbf{A(a,}0\mathbf{,}0)$ (\ref{qam}). It means that the
conformal limit of our SM is dual to CFT (\ref{ac1}).

Using the amplitudes $\mathbf{A(a,0,}0)$ (\ref{qam}) we can write the
equations for quantization of $P.$ This procedure is explained in details in
\cite{BK}. In semiclassical limit they have the form%
\begin{equation}
2unP+\sum_{\alpha>0}e_{\alpha}f_{1}(e_{\alpha}\cdot P)+n\sum_{j=1}^{n}%
h_{j}f_{2}(h_{j}\cdot P)=2\pi\left(  \rho+\Omega_{n}\right)  ,
\end{equation}
where $f_{1}(z)=\frac{1}{\mathrm{i}}\log\left(  \frac{\Gamma(1+\mathrm{i}%
z)}{\Gamma(1-\mathrm{i}z)}\right)  ,f_{2}(z)=\frac{1}{\mathrm{i}}\log\left(
\frac{\Gamma(\frac{1}{2}-\mathrm{i}z)}{\Gamma(\frac{1}{2}+\mathrm{i}%
z)}\right)  ,$ and $\Omega_{n}$ is the highest weight of $su(n)$.

In the deep UV region $u>>1,$ we can expand this equation up to the first
order in $P.$ We have $\Sigma_{\alpha>0}(e_{\alpha})_{s}(e_{\alpha}%
)_{r}=n\delta_{s,r},\Sigma_{j}^{n}(h_{j})_{s}(h_{j})_{r}=\delta_{s,r}$ and
$\frac{\Gamma^{\prime}(1)}{\Gamma(1)}-\frac{\Gamma^{\prime}(\frac{1}{2}%
)}{\Gamma(\frac{1}{2})}=\log4.$ Then we derive the semiclassical quantization
of $P,$%
\[
P=\frac{\pi\left(  \rho+\Omega_{n}\right)  }{n(u+n\log4)}.
\]
For $\mathrm{e}_{0}=12P^{2}=p(c_{UV}-c_{\rm eff}(R)  ),$ we obtain that for
$u>>1$ one has%
\begin{equation}
\mathrm{e}_{0}(u)=\frac{12\pi^{2}\rho^{2}}{n^{2}(u+n\log4)^{2}}+O(1/u^{5}%
)=\frac{\pi^{2}(n^{2}-1)}{n(u+n\log4)^{2}}+O(1/u^{5}). \label{eouv}%
\end{equation}
It implies that $\mathrm{e}_{0}(u)=\frac{3n}{2u}(1+O(u^{4}))$ for $u<<1$ and
$\mathrm{e}_{0}(u)=\frac{\pi^{2}(n^{2}-1)}{n(u+n\log4)^{2}}+O(1/u^{5})$ for
$u>>1.$ Between these values, $\mathrm{e}_{0}(u)$ is monotonically decreasing
(due to $c-$theorem \cite{Z}) function of $u.$ We note that $p=1+b^{2}\eqsim
b^{2}>>1.$ It means that%
\begin{equation}
c_{\rm eff}(R)-c_{UV}\eqsim-\frac{3n}{2\log\left(  \frac{1}{MR}\right)  }\ \ ;\quad
c_{\rm eff}(R)-c_{UV}\eqsim-\frac{\pi^{2}(n^{2}-1)b^{2}}{n(\log\left(  \frac{1}%
{MR}\right)  +b^{2}n\log4)^{2}}%
\label{cRRR}
\end{equation}
for $b^{2}>>\log\left(  \frac{1}{MR}\right)  $ and $b^{2}<<\log\left(
\frac{1}{MR}\right)  .$

The exited levels $\mathrm{e}_{i}(u)-\mathrm{e}_{0}(u)$ in the
\textquotedblleft isotopic symmetric\textquotedblright\ sector depend on $n-1$
quantum numbers and flow from $\lambda_{i}/u,$ where $\lambda_{i}$ are the
eigenvalues of Laplacian on $CP(n-1)$ for small $u,$ to $\frac{12\pi
^{2}\left(  \Omega_{n}^{2}+2\rho\cdot\Omega_{n}\right)  }{n^{2}(u+n\log4)^{2}%
}$ for large $u.$

\section{Integrable perturbation $\mu_{1}e^{b(e_{0}\cdot\varphi)}$}

In this Section we briefly consider the second integrable perturbation of CFT
(\ref{ac1}). With this perturbation the action will have the following form%
\begin{equation}
\mathcal{A}_{sl(n)}^{(2)}\mathcal{=A}_{C}+\mu_{1}\int d^{2}xe^{b(e_{0}%
\cdot\varphi)}, \label{act2}%
\end{equation}
where $e_{0}$ is affine root of $sl(n)$.

We can use again the Coleman-Mandelstam $2d$ correspondence between fermion
and bosons (\ref{CM}). We introduce the fields%
\[
\xi_{i}=a(\eta_{i}\cdot\vartheta)+\vartheta_{0}\frac{\sqrt{n+b^{2}}}%
{\sqrt{n(n-1}};\quad(\eta_{i}\cdot\eta_{j})=1-\frac{\delta_{i,j}}{n-1};\quad
i=1,..n-1
\]
and projectors $\gamma_{+},\gamma_{-}:\gamma_{\pm}=\frac{1}{2}(1\pm\gamma
_{5}).$ Then the action (\ref{act2}) after application of Coleman-Mandelstam
rules can be rewritten in the form suitable for perturbation theory for small
$b.$%
\begin{equation}
\mathcal{A}_{sl(n)}^{(2)}=\int d^{2}x\left(  L_{F}^{(2)}+L_{B}^{(2)}%
+L_{FB}^{(2)}\right)  \label{L(2)}%
\end{equation}
where $L_{F}^{(2)}$ and $L_{FB}^{(2)}$ can be derived from $L_{F}$ and
$L_{FB}$ in (\ref{acf}) by taking the sums from $1$ to $n-1$ and substitution
$g_{1}\rightarrow g_{1}^{(2)}=-\frac{b^{2}(n-1+b^{2})}{(1+b^{2})(n+b^{2}%
)},g_{2}\rightarrow g_{2}^{\left(  2\right)  }=\frac{b^{2}}{(n+b^{2}%
)(1+b^{2})}.$

This QFT has a broken $\mathbf{P}$ symmetry ($n>2$)$\ $but possesses
$\mathbf{PT}$ and $\mathbf{C}$ symmetries. For small $b$ it has $2(n-1)$
particles $\psi_{i},\psi_{i}^{\ast}$ with mass $m,$ and $n-1$ scalar particles
$\varphi_{i}$ which can be considered as bound states of Fermi particles with
masses $m_{j}=2m\sin\left(  \frac{\pi j}{n}\right)  +O(b^{2}).$ For finite $b$
they disappear from the spectrum.

The scattering theory now is described by the eqs.\ref{S} where in $S$- matrix
$S^{(2)}(\theta)$ the indices $i\rightarrow\psi_{i},\overline{i}%
\rightarrow\psi_{i}^{\ast}$ now take values $1,..,n-1.$ In the Eqs.(\ref{S})
we also take the parameters $n\rightarrow\mathsf{n=n(b}^{2})$ and
$\lambda=\lambda(b^{2}).$ The unitarizing factor $F\left(  \theta\right)  $
now becomes%
\begin{equation}
F_{\mathsf{n,\lambda}}^{(2)}\left(  \theta\right)  =\exp\left(  \mathrm{i}%
{\displaystyle\int\limits_{-\infty}^{\infty}}\frac{d\omega}{\omega}\,
\frac{\sinh\left(  \frac{\pi(\mathsf{n}-1)\omega}{\mathsf{n}}\right)
\sinh\left(  \frac{\pi(p-1)\omega}{\mathsf{n}}\right)  }
{\sinh(\pi\omega)\sinh\left(  \frac{\pi p\omega}{\mathsf{n}}\right)  }\sin(\omega
\theta)\right)  ;\quad\lambda=\frac{1}{p}. \label{F2}%
\end{equation}
The dependence of the parameters $\mathsf{n(b}^{2})$ and $\lambda(b^{2})$ can
be derived from BA equations for GSE in external fields $A_{i}$ coupled with
$n-1$ conserved $U(1)$ charges $Q_{i}=\int\psi_{i}^{\ast}\psi d^{2}x.$
Following the lines of Section 5, we find that%
\begin{equation}
\frac{1}{\lambda}=p=(1+b^{2});\quad\mathsf{n=(}n\mathsf{+}b^{2}). \label{nbr}%
\end{equation}

At the limit $b^{2}\rightarrow\infty,$the scattering matrix $S^{(2)}(\theta)$
tends to the identity matrix and can be expanded in $1/b^{2}=\mathrm{b}^{2}.$
It is important to note that the function $\ F_{\mathsf{n,\lambda}}%
^{(2)}\left(  \theta\right)  $ (\ref{F2}) at small $b^{2}$ is $1+O(b^{2})$ and
amplitudes $\psi_{i}\psi_{i}\rightarrow\psi_{i}\psi_{i}$ equal to
$-F_{\mathsf{n,\lambda}}^{(2)}\left(  \theta\right)  \rightarrow-1$ at
$\theta\rightarrow0,$ what is characteristic property for fermionic particles,
for $b^{2}\rightarrow\infty$ we get that this amplitude $-F_{\mathsf{n,\lambda
}}^{(2)}\left(  \theta\right)  \rightarrow1+O(\mathrm{b}^{2}),$ as it happens
for bosonic particles. This phenomenon (fermion-boson duality) is easy to see
in the case $n=2$ \ where we have only one charged particle $\psi,\psi^{\ast
},$ and hence, $S$-matrix is a pure phase. The action (\ref{L(2)}) is
described by $L_{sl(2)}^{\left(  2\right)  }$, which is:%
\[
\frac{1}{8\pi}\left(  \partial_{\mu}\varphi)^{2}+2(\mathrm{i}\overline{\psi
}\gamma_{\mu}\partial_{\mu}\psi+m\overline{\psi\psi}e^{b\varphi/\sqrt{2}%
}-\frac{b^{2}}{2+b^{2}}(\overline{\psi}\gamma_{\mu}\psi)^{2})+m^{2}\cosh
(\sqrt{2}b\varphi)\right)
\]
The scattering matrix $S_{\psi\psi}=-F_{\mathsf{n,\lambda}}^{(2)}\left(
\theta\right)  $ for $n=2$ can be calculated and is given by%
\begin{equation}
S_{\psi\psi}(\theta)=-\frac{\cosh(\theta/2+\mathrm{i}\Delta/2)}{\cosh
(\theta/2-\mathrm{i}\Delta/2)},
\qquad
S_{\psi\psi^{\ast}}=S_{\psi\psi
}(\mathrm{i}\pi-\theta),
\qquad
\Delta=\frac{\pi b^{2}}{1+b^{2}}. \label{CsG}%
\end{equation}
For $b>>1$ it can be expanded as a regular series in $1/b^{2}=\mathrm{b}^{2}.$
In this region our QFT can be described in terms of one complex scalar field
$\omega(x)$ by the action \cite{F3}%
\begin{equation}
\mathcal{A}_{sl(2)}^{(2)}\mathcal{=}\int d^{2}x\frac{1}{4\pi}\left(
\frac{\partial_{\mu}\omega\partial_{\mu}\omega^{\ast}}{1+\mathrm{b}^{2}%
\omega\omega^{\ast}}+m^{2}\omega\omega^{\ast}\right)  . \label{SMp}%
\end{equation}
We see that charged particles being fermions in the weak coupling region,
become bosons in the strong coupling region.

The SM part of the action (\ref{SMp}) coincides with the dual action for CFT
(\ref{ac1}) $\left(  n=2\right)  $, or with the conformal limit of metric of
deformed $CP\left(  1\right)  $ model. The potential part $m^{2}\omega
\omega^{\ast}$ defines the tachyon. \ In general case (arbitrary $n$) bosonic
QFTs dual to (\ref{L(2)}) for small $\mathrm{b}$ have the classical limit.
They define classical integrable models, which we call conventually
\textquotedblleft non-abelian $sl\left(  n\right)  $ affine Toda
theories\textquotedblright. \

The SM part of these QFTs is described by the conformal limit of the metric
$G_{ij}^{\left(  c\right)  }$ and the $B^{(c)}-$field of deformed $CP\left(
n-1\right)  $ model. Together with potential (tachyon) the dual action can be
expressed in terms of $n-1$ complex fields $\omega_{i}(x).$%
\begin{equation}
\mathcal{A}_{sl\left(  n\right)  }^{(2)}=\int\frac{d^{2}x}{4\pi}\left(
G_{ij}^{\left(  c\right)  }(\mathrm{b}\omega_{i},\mathrm{b}\omega_{i}^{\ast
})(\partial_{\mu}\omega_{i}\partial_{\mu}\omega_{j}^{\ast}+\varepsilon
_{\alpha\beta}B_{ij}^{\left(  c\right)  }\partial_{\alpha}\omega_{i}%
\partial_{\beta}\omega_{j}^{\ast})+\frac{m^{2}}{\mathrm{b}^{2}}\mathsf{T}%
\right)  , \label{a2}%
\end{equation}
where $\mathsf{T}\mathcal{=}\frac{m^{2}}{\mathrm{b}^{2}}\mathsf{T}%
(\mathrm{b}\omega_{i},\mathrm{b}\omega_{i}^{\ast})$ denotes the tachyon.
Contrary to the scattering theory and observables, the functions
$G_{ij}(\omega_{i},\omega_{i}^{\ast}),B_{ij}(\omega_{i},\omega_{i}^{\ast})$
and $\mathsf{T}(\omega_{i},\omega_{i}^{\ast})$ can depend on the choice of
coordinates (see appendix \ref{ap2}). We hope to return to this problem in a
separate publication.

In the end of this section we make some important remark, not related to the
duality. In the action $\mathcal{A}_{sl\left(  n\right)  }^{(2)},$ which has a
classical limit, we can take parameter $\mathrm{b}$ purely imaginary:
$\mathrm{b\rightarrow i\gamma.}$ The classical theory will be integrable and
well-defined. However, due to the singularities of the metric and the $B$-field, 
the quantization of this theory will be more subtle.
For example, the classical theory (\ref{SMp}) after $\mathrm{b\rightarrow
i\gamma}$ is known as complex Sine-Gorgon or Lund-Regge \cite{LR} model. In
quantum case we can perform in $S$-matrix (\ref{CsG}) the transformations
$b\rightarrow\frac{1}{\mathrm{b}},\mathrm{b\rightarrow i\gamma,}$ and denote
$S_{\psi\psi}(\theta)\rightarrow S_{\omega\omega}(\theta).$%
\[
S_{\omega\omega}(\theta)=-\frac{\sinh(\theta/2+\mathrm{i}\Delta_{1}/2)}%
{\sinh(\theta/2-\mathrm{i}\Delta_{1}/2)};\quad S_{\omega\omega^{\ast}}\left(
\theta\right)  =S_{\omega\omega}(\mathrm{i}\pi-\theta),\quad\Delta_{1}%
=\frac{\pi\gamma^{2}}{1-\gamma^{2}}.
\]
These amplitudes are in exact agreement with perturbation theory in $\gamma.$
However, the amplitude $S_{\omega\omega}(\theta)$ possesses the pole in the
physical strip, corresponding to a bound state of two particles $\omega
(\theta).$ This particle has ~$U\left(  1\right)  $ charge equal twice the
charge of $\omega(\theta).$ For general values of $\gamma$ the $S$-matrix
bootstrap is not closed and the charge of particles is not limited. This
contradiction disappears only for special values of $\gamma,$ namely when%
\begin{equation}
\Delta_{1}=\frac{\pi\gamma^{2}}{1-\gamma^{2}}=\frac{\pi}{k},\quad\gamma
^{2}=\frac{1}{k+n},\quad k=0,2,3.. \label{qcc}%
\end{equation}
It means that we have a well-defined QFT only for special values of $\gamma$
(quantization of coupling constant)$.$ The similar phenomenon has place for all
values of $n.$ The analysis of scattering theory shows that theory is
well-defined only for $\gamma^{2}=\frac{1}{k+n}$ with positive integer $k$.\footnote{The
similar phenomenon appears in the non-unitarian imaginary Toda QFT which is
described by the Lagrangian $L_{B}$ (\ref{acf}) with purely imaginary
$b=\mathrm{i}\mathsf{b}$. In this QFT the scattering bootstrap is closed only for
$\mathsf{b}^{2}=\frac{p}{p+1},$ where $p$ is integer and $p>h$. For these values
of the coupling constant the imaginary Toda theory, after the quantum group
restriction, becomes a unitary QFT.}
The condition%
\begin{equation}
\frac{1}{\gamma^{2}}=-b^{2}=k+n \label{end}%
\end{equation}
will be considered in the next section.

\section{Rational CFTs, their perturbations and sigma models}

In this Section we discuss some important points (in particular, another
integrable deformations of SMs) which were not considered in the main body of
this paper.

\bigskip

$\mathit{1.}$ \ In section 6 we discussed the SMs and Ricci flows related with
massive QFTs. However there are massless Ricci flows associated with SMs with
singular metrics which for special values of the parameter $b$ describe the
flows from CFTs with continuous spectrum in the UV regime to rational
unitarian CFTs in IR regime (See \cite{FO},\cite{FS},\cite{FL}). For deformed
$CP\left(  n-1\right)  $\ the SM representation dual to CFT (\ref{ac1}) is
described by the action (\ref{a2}) with $b^{2}>0$ and $T=0.$ The \ Ricci flow
with singular metric relates CFT (\ref{a2}) with $b^{2}=n+k$ \ and rational
CFT with $-b^{2}=n+k$ (\ref{end}).

The central charge of CFT (\ref{ac1}) with $b^{2}=-(n+k),a^{2}=-(n+k-1)$ is
(\ref{cn})%
\begin{equation}
\mathrm{c}_{n,k}=\left(  n-1\right)  \left(  2-\frac{(n+1)n}{k+n}%
+\frac{n(n-2)}{k+n-1}\right)  \label{cnk1}%
\end{equation}
here we ignore for the moment $1$ coming from free field $\vartheta_{n}.$ This
field will appear later.

The central charge (\ref{cnk1}) corresponds to the coset $\frac{SU(n)_{k}%
}{SU(n-1)_{k}U(1)}.$ These minimal models $\mathcal{M}\left(  n,k\right)  $
are characterized by strongly degenerate representations of $W-$algebras
$\mathcal{W}\left(  n,k\right)  $ which commute with all screening charges
with densities $\left\{  V_{1,}V_{-2}\right\}  ,..,\left\{  V_{n-1,}%
V_{-n}\right\}  $ (\ref{cpai}) and can be represented by $2\left(  n-1\right)
$ free fields $(n-1)$ dimensional vector $\varphi,\left(  n-2\right)  $
dimensional $\vartheta$ and field $\vartheta_{0}.$ The primary fields in
$\mathcal{M}\left(  n,k\right)  $ can be represented as%
\[
\mathbf{\Phi}\left(  \mathbf{\Omega}_{n},\mathbf{\Omega}_{n-1},s\right)
=\exp\left(  \mathrm{i}\frac{\mathbf{\Omega}_{n}\cdot\varphi}{\sqrt{n+k}%
}-\frac{\mathbf{\Omega}_{n-1}\cdot\vartheta}{\sqrt{n+k-1}}-\frac{\vartheta
_{0}s}{\sqrt{kn\left(  n-1\right)  }}\right)
\]
where $\mathbf{\Omega}_{n}$ is the highest weight vector of $su(n)$ and $\pm
s$ is integer with the condition%
\[
k\geq-\mathbf{\Omega}_{n}\cdot(\mathbf{e}_{0})_{n}\geq-\mathbf{\Omega}%
_{n-1}\cdot(\mathbf{e}_{0})_{n-1}\geq|s|
\]
The dimensions of these fields
\begin{equation}
\mathbf{\Delta}(\mathbf{\Omega}_{n},\mathbf{\Omega}_{n-1},s)  =
\frac{1}{2}\left(  \frac{\mathbf{\Omega}_{n}\cdot\left(  \mathbf{\Omega}%
_{n}+2\rho_{n}\right)  }{n+k}-\frac{\mathbf{\Omega}_{n-1}\cdot\left(
\mathbf{\Omega}_{n-1}+2\rho_{n-1}\right)  }{n+k-1}-\frac{s^{2}}{kn\left(
n-1\right)  }\right)  .
\end{equation}
In particular, the receiving field of Ricci flow is the descendent field
$\mathbf{\Phi}_{R}=(\mathcal{W}_{3})_{-1}\mathbf{\Phi}\left(  \mathbf{ad},\mathbf{0},0\right)  $ with dimension $1+\frac{n}{k+n}.$

For many applications it is useful to use the level-rank duality \cite{ALT}
which relates $\frac{SU(n)_{k}}{SU(n-1)_{k}U(1)}$ and $\frac{SU(k)_{n-1}%
SU(k)_{1}}{SU(k)_{n}}$ minimal models. The minimal models $\mathcal{M}\left(
k,n\right)  =\frac{SU(k)_{n-1}SU(k)_{1}}{SU(k)_{n}}$ are well known. They
possess $W_{k}-$symmetry associated with the Lie algebra $su(k)$ and central
charge $c_{k,n}=(k-1)\left(  1-\frac{k(k+1)}{(k+n-1)(k+n)}\right)  .$ It is
easy to see that $c_{k,n}=\mathrm{c}_{n,k}.$ The space of primary fields
$\Phi_{k}\left(  \Omega,\Omega^{\prime}\right)  \in\mathcal{M}\left(
k,n\right)  $ is specified by two highest weight vectors of $su(k)$: $\Omega$ and $\Omega^{\prime}$ which satisfy the conditions: 
$-\Omega\cdot e_{0}\leq n,$ and $-\Omega^{\prime}\cdot e_{0}\leq n-1$ and
have the dimensions
\begin{equation}
\Delta\left(  \Omega,\Omega^{\prime}\right)  =\frac{1}{2}\left(  \frac{a}%
{b}\Omega-\frac{b}{a}\Omega^{\prime}\right)  \left(  \frac{a}{b}\Omega
-\frac{b}{a}\Omega^{\prime}+\frac{2}{ab}\rho_{k}\right)  ,
\end{equation}
where $b=\textrm{i}\sqrt{k+n}$, $a=\textrm{i}\sqrt{k+n-1}.$ In particular the
receiving field of Ricci flow is the descendent field $\Phi_{R}=(W_{3}%
)_{-1}\Phi_{k}\left(  ad,0\right)  $ with dimension $1+\frac{n}{k+n}.$

All currents of $W_{k}-$algebra (which we call for fixed $n$ as $W_{k}\left(
n\right)  $) can be represented by $ k-1$ free fields $\chi$
and commute with screenings generated by fields $e^{\mathrm{i}\frac{a}{b}%
e_{j}\cdot\chi},e^{-\mathrm{i}\frac{b}{a}e_{j}\cdot\chi}.$ The fields
$\Phi_{k}\left(  \Omega,\Omega^{\prime}\right)  $ can be represented as%
\begin{equation}
\Phi_{k}\left(  \Omega,\Omega^{\prime}\right)  =\exp\left(  \mathrm{i}\frac
{a}{b}\Omega\cdot\chi-\mathrm{i}\frac{b}{a}\Omega^{\prime}\cdot\chi\right)
\end{equation}
It means that we have two different representations for the same CFT
$\mathcal{M}(  k,n)  =\mathcal{M}(  n,k)$. We call
$W_{k}(  n)$ as \textquotedblleft vertical\textquotedblright%
\ $W-$algebra and $\mathcal{W}\left(  n,k\right)  $ as \textquotedblleft
horizontal\textquotedblright\ $W-$algebra of CFT $\mathcal{M}(k,n)$. The existence of two different representations for the same CFT
and as a result two different integral representations for the correlation
functions will be studied in a separate publication. Here we give the simplest
and known example of horizontal $W-$algebra.

Such example is provided by algebra $\mathcal{W}(2,k)$, which
for integer $k$ is the symmetry algebra of CFT $\mathcal{M}(k,2)
$ with central charge $c_{k,2}=\mathrm{c}_{2,k}$ $=\frac{2(k-1)}{k+2}.$ These
models possess the symmetry generated by $su(2)$ or $Z_{k}$ parafermions
$\Psi_{i}(z),\Psi_{i}^{\ast}=\Psi_{k-i}$ \cite{ZF} with spin $\frac{i(k-i)}%
{k}.$ All parafermions and all symmetry algebra can be derived from $\Psi_{1}$
which has a representation by two fields $\varphi,\vartheta_{0}$ \cite{N}$:$%
\[
\Psi_{1}=\left(  \sqrt{\frac{k+2}{k}}\mathrm{i}\partial\varphi+\partial
\vartheta_{0}\right)  e^{\sqrt{\frac{2}{k}}\vartheta_{0}},
\qquad
\Psi_{1}^{\ast}=\left(  \sqrt{\frac{k+2}{k}}\mathrm{i}\partial\varphi-\partial\vartheta
_{0}\right)  e^{-\sqrt{\frac{2}{k}}\vartheta_{0}}%
\]
The fields $\Psi_{1}(z),\Psi_{1}^{\ast}(z)$ commute with screening charges
generated by $\left\{  V_{1,}V_{-2}\right\}  ,$ which for $b^{2}=-(n+k)$ are%
\begin{equation}
V_{1}=\exp\left(  \mathrm{i}\sqrt{\frac{k+2}{2}}\varphi+\sqrt{\frac{k}{2}%
}\vartheta_{0}\right)  ,
\qquad
V_{-2}=\exp\left(  \mathrm{i}\sqrt{\frac{k+2}{2}}\varphi-\sqrt{\frac{k}{2}}\vartheta_{0}\right)  \label{Vp}%
\end{equation}
Hence all fields appearing in OPE $\Psi_{1}(z)\Psi_{1}^{\ast}(z^{\prime})$
commute with screenings and form $\mathcal{W}\left(  2,k\right)  .$ Explicitly
we have:%
\begin{align}
\left(  z^{\prime}-z\right)  ^{2\Delta_{1}}\Psi_{1}(z^{\prime})\Psi_{1}^{\ast
}(z)  &  =I+\left(  z^{\prime}-z\right)  ^{2}\frac{k+2}{k}T\left(  z\right)
\nonumber\\
&  +\left(  z^{\prime}-z\right)  ^{3}\left(  \mathcal{W}_{3}(z)+\frac
{T^{\prime}(z)}{2}+\right)  +\cdots, \label{par}%
\end{align}
where $T(z)=-\frac{1}{2}\left(  (\partial\varphi)^{2}+(\partial\phi
)^{2}\right)  +\frac{\mathrm{i}}{\sqrt{2(k+2)}}\partial^{2}\varphi$ and the
local current $\mathcal{W}_{3}(z)$ generating $\mathcal{W}\left(  2,k\right)
$ algebra up to numerical factor is%
\begin{align}
\mathcal{W}_{3}(z)  &  =\partial\phi\left(  \sqrt{\frac{2}{9}}\left(
3k+4\right)  (\partial\phi)^{2}+\sqrt{2}(k+2)(\partial\varphi)^{2}%
+\mathrm{i}\left(  2+k\right)  ^{3/2}\partial^{2}\varphi\right)  \nonumber\\
&  +\mathrm{i}\partial^{2}\phi\partial\varphi\left(  k\left(  2+k\right)
^{1/2}\right)  +\partial^{3}\phi\frac{1}{3\sqrt{2}} \label{w32}%
\end{align}
More complicated field $\mathcal{W}_{3}(z)$ generating $\mathcal{W}\left(
3,k\right)  $ is given in appendix \ref{ap1}.

\bigskip

$\mathit{2.}$ \ The CFTs $\mathcal{M}\left(  k,n\right)  $ and $\mathcal{M}%
\left(  n,k\right)  $ describe the critical points of RG flows. It is
interesting to consider the integrable and non-integrable QFTs corresponding
to perturbed $\mathcal{M}\left(  k,n\right)  $ $(\mathcal{M}\left(
n,k\right)  )$ CFTs, in particular, the perturbation considered at the end of
section 8 (the tachyon $T$) with $\gamma$ given by (\ref{end}). In this case
we derive the integrable QFTs corresponding to perturbation operator $\Phi
_{k}(ad,0)$ $\left(  \mathbf{\Phi}\left(  \mathbf{ad},\mathbf{0},0\right)
\right)  $ with dimension $\Delta_{ad}=\frac{n}{k+n}.$ The scattering theory
for these QFTs was constructed in \cite{HDV}.

More interesting are the perturbations corresponding to deformed
$CP(n-1)-$sigma models and the same models with fermion. First we consider the
model which is non-integrable for $n>2,$ but for $k\gg n$ has all
characteristic properties of $CP(n-1)-$sigma model without fermion. We
introduce the perturbing operators $\Lambda_{f}$ and $\Lambda_{f}^{\ast}$, 
where fields $\Lambda_{f}=$ $\Phi_{k}(0,\omega_{1})$ and
$\Lambda_{f}^{\ast}=$ $\Phi_{k}(0,\omega_{k-1})$, associated with the vector representations of $su(k)$,
have left and right
dimensions $\Delta_{f}=1-\frac{(k+1)n}{2k(k+n-1)}$ and the theory with action%
\begin{equation}
\mathcal{A}_{\theta}\mathcal{=A}_{CFT}+\lambda\int(e^{\mathrm{i}\theta\pi
/k}\Lambda_{f}+e^{-\mathrm{i}\theta\pi/k}\Lambda_{f}^{\ast})d^{2}x
\label{actop}%
\end{equation}
For $n=2,$ $\Lambda_{f}=\Psi_{1}\overline{\Psi}_{1},\Lambda
_{f}^{\ast}=\Psi_{1}^{\ast}\overline{\Psi}_{1}^{\ast}$ and the theory is
integrable \cite{F4},\cite{FAL} for $\theta=0$ and $\theta=$ $\pi$. In the
limit $k\rightarrow\infty,$ it coincides with $O(3)$ or $CP(1)-$sigma model
with parameter $\theta,$ coupled with topological charge as \textrm{i}$\theta
T$. It was shown that for $\theta=0$ the theory is massive and for $\theta=$
$\pi$ massless. The effective central charge of the theory living on the
finite space circle of length $R,$ flows from $2$ (in UV) to $0$ (in IR) for
$\theta=0$ and from $2$ to $1$ for $\theta=$ $\pi.$ The crucial role in this
behavior is played by the instanton contributions. Similar contributions appear for $n>2.$ The effective central
charge $c_{\rm eff}(R)$ has two types of contributions. The perturbative
contributions and \textquotedblleft instanton\textquotedblright\ ones. The
first come in the perturbed CFT from the correlation functions%
\begin{equation}
\lambda^{2m}\left\langle \Lambda_{f}(x_{1})\Lambda_{f}^{\ast}(x_{2}%
),\dots,\Lambda_{f}(x_{2m-1})\Lambda_{f}^{\ast}(x_{2m})\right\rangle \label{cf1}%
\end{equation}
and the second from the correlation functions%
\begin{equation}
e^{\mathrm{i}\theta\pi q}\lambda^{2m+kq}\left\langle \Lambda_{f}(y_{1}%
)\Lambda_{f}(y_{2})..\Lambda_{f}(y_{kq});\Lambda_{f}(x_{1})\Lambda_{f}^{\ast
}(x_{2}),..,\Lambda_{f}(x_{2m-1})\Lambda_{f}^{\ast}(x_{2m})\right\rangle
\label{cf2}%
\end{equation}
These correlation functions do not vanish because in CFTs $\mathcal{M}(k,n)$ 
the fields $\Lambda_{f}$ and $\Lambda_{f}^{\ast}$ have $Z_{k}$
charge $1$ and $-1$. As a result the effective central charge can be expanded
as%
\[
c_{\rm eff}^{(\theta)}(  R)  =c_{n,k}+%
{\displaystyle\sum\limits_{q=0,m=0}^{\infty}}
\cos(\pi\theta q)d_{m}^{(q)}\left(  M_{s}R\right)  ^{(2kq+4m)\left(
1-\Delta_{f}\right)  }%
\]
where $1-\Delta_{f}=\frac{(1+k)n}{2k(k+n-1)}=\frac{n}{2k}(1+O(1/k));$
$c_{n,k}=2\left(  n-1\right)  \left(  1+O(1/k)\right)  $ and mass parameter
$M_{s}\thicksim\lambda^{\frac{1}{2(1-\Delta_{f})}}.$ All coefficients
$d_{m}^{(q)}$ can be derived by integration of correlation functions
(\ref{cf1},\ref{cf2}). These series are very similar to those for $n=2.$ For
example, the first instanton contribution (up to the factor $d_{0}^{(1)}%
,$which can be calculated exactly) is proportional $\left(  M_{s}R\right)
^{n},$ in agreement with renormalization group for $CP(n-1)$ QFT. It is
natural to expect that for $\theta=0$ the QFT (\ref{actop}) will be massive
and for $\theta=\pi$ it will be massless.

To derive the integrable QFT describing in the limit $k\rightarrow\infty$,
$CP(n-1)-$model with fermion, we take into account the field $\vartheta_{n}$
which was completely irrelevant in the action $\mathcal{A}_{CFT}$ (\ref{ac1}).
The field $\vartheta_{n}(x)=\theta(x)$ will play a role of fluctuating
topological parameter $\theta.$ We introduce the free massless field
$\theta(x)$ and modify the fields $(\Lambda_{f}(x),\Lambda_{f}^{\ast}%
(x))\in\mathcal{M}\left(  k,n\right)  $ as%
\[
\Lambda=\Lambda_{f}\exp\left(  \mathrm{i}\nu_{n,k}\theta(x)\right),
\qquad
\Lambda^{\ast}=\Lambda_{f}^{\ast}\exp\left(  -\mathrm{i}\nu_{n,k}%
\theta(x)\right),
\qquad
\nu_{n,k}^{2}=\left(  \frac{(n+k)}{k(k+n-1)}\right).
\]
The fields $(\Lambda(x),\Lambda^{\ast}(x))$ have the right and left dimensions
$\Delta=1-\frac{n-1}{2(n+k-1)}.$ The QFT with action%
\begin{equation}
\mathcal{A}_{\theta(x)}\mathcal{=A}_{CFT}+\int\left(  \frac{\partial_{\mu
}\theta(x)\partial_{\mu}\theta(x)}{8\pi}+\varkappa(\Lambda+\Lambda^{\ast
})\right)  d^{2}x \label{acfer}%
\end{equation}
is integrable. Its scattering theory can be derived by the quantum group
restriction from the $S$-matrix described in section 4. With this scattering
theory we can write the TBA equations for effective central charge. The
massless field $\theta(x)$ cancels all \textquotedblleft
instanton\textquotedblright\ contributions, and the series for $c_{\rm eff}(R)$
simplify%
\[
c_{\rm eff}\left(  R\right)  =c_{n,k}+1+%
{\displaystyle\sum\limits_{m=2}^{\infty}}
\mathrm{d}_{m}\left(  MR\right)  ^{2\frac{m(n-1)}{k+n-1}}.
\]
The physical mass $M$ is related with coupling constant $\varkappa$ as follows%
\[
\frac{\pi\varkappa\Gamma\left(  \frac{1}{n+k-1}\right)  }{\Gamma\left(
1-\frac{1}{n+k-1}\right)  k}=\left(  \frac{n\left(  n-1\right)  }{2}\right)
^{1/2}\left(  M\frac{\Gamma\left(  \frac{1}{n}\right)  \Gamma\left(
\frac{n+k-1}{n}\right)  }{4\Gamma\left(  \frac{k}{n}\right)  }\right)
^{\frac{\left(  n-1\right)  }{(n+k-1)}}%
\]
Using this relation and properties of the coefficients $\mathrm{d}_{m},$ it is
possible to calculate the scaling limit $(k\gg1,\log1/MR)\gg1.\frac
{\log(1/MR)}{k}$ fixed$)$ of the effective central charge%
\begin{equation}
c_{\rm eff}(  R)  =2n-1+\frac{3n\left(  n-1\right)  }{k}\left(  1-\left(
MR\right)  ^{\frac{2\left(  n-1\right)  }{(n+k-1)}}\right)  ^{-1}.
\label{ceffR}
\end{equation}
In the limit $k\rightarrow\infty,$ $c_{\rm eff}(R)  =1+2(n-1)-\frac
{3n}{2\log(1/MR)}.$ This result coincides with RG equation for $CP(n-1)$ model
with fermion.

We note that for constant $\theta(x)=\theta\frac{\pi}{k\nu}$ the QFTs
(\ref{acfer}) and (\ref{actop}) coincide. If we consider the action of
$CP(n-1)-$model with fermion%
\[
\mathcal{A}_{CP(n-1)+\chi}=\mathcal{A}_{CP\left(  n-1\right)  }+\frac{1}{4\pi
}\int d^{2}x\left(  \mathrm{i}\overline{\chi}\gamma_{\mu}(\partial_{\mu
}+B_{\mu})\chi\right)
\]
and use the relations \textrm{i}$\overline{\chi}\gamma_{\mu}\partial_{\mu}%
\chi=\frac{1}{2}\partial_{\mu}\theta(x)\partial_{\mu}\theta(x)$ and
$\overline{\chi}\gamma_{\mu}\chi=\varepsilon_{\mu\sigma}\partial_{\sigma
}\theta(x)$ then after the integration by parts with identity $\varepsilon
_{\mu\sigma}\partial_{\sigma}B_{\mu}=4\pi T(x),$ where $T(x)$ is the density
of topological charge we derive%
\begin{equation}
\mathcal{A}_{CP(n-1)+\chi}=\mathcal{A}_{CP\left(  n-1\right)  }+\int
d^{2}x\left(  \frac{\partial_{\mu}\theta(x)\partial_{\mu}\theta(x)}{8\pi
}+\mathrm{i}\theta(x)T(x)\right)  . \label{aaa}%
\end{equation}
We see that field $\theta(x)$ plays the role of axion \cite{B}. For constant
$\theta(x)$ this action coincides with the action of $CP(n-1)-$model with
topological term. It has very rich dynamics, but does not \ define the
integrable QFT .

\bigskip

$\mathit{3.}$ \ Studying the deformed SMs on the groups and on the symmetric
spaces it is useful to consider the extended set of perturbed rational CFTs.
Here we consider the well known rational coset CFTs $\mathcal{M}\left(
G,m,l\right)  =\frac{G_{m}G_{l}}{G\left(  m+l\right)  },$ where $G$ is the
simply laced Lie algebra, perturbed by integrable field $\Phi$ associated with
the adjoint representation of $G$ and conformal dimension $\Delta=1-\frac
{h}{m+l+h}.$ The central charge of $\mathcal{M}\left(  G,m,l\right)  $ is%
\begin{equation}
c\left(  G,m,l\right)  =\frac{r\left(  h+1\right)  lm\left(  l+m+2h\right)
}{\left(  l+m+h\right)  \left(  m+h\right)  \left(  l+h\right)  } \label{cG}%
\end{equation}
where $r$ and $h$ are rank and Coxeter number of $G,D=r\left(  h+1\right)
=\dim\left(  G\right)  .$

The action of integrable perturbed CFT $\mathcal{M}\left(  G,m,l\right)  $ is%
\begin{equation}
\mathcal{A}_{G}=\mathcal{A}_{G}\left(  CFT\right)  -\kappa\int\Phi\left(
x\right)  d^{2}x\label{aGml}%
\end{equation}
For $\kappa>0$ these QFTs are massive, the particles for finite $m,l$ are the
kinks\footnote{The scattering matrix of the basic kinks is derived from that for deformed
principle chiral field (see footnote 3) equal to $S(G)_{p_{1},p_{2}}\left(
\theta\right)  ,$ where $p_{1}=h+m$, $p_{2}=h+l$, by the quantum group restriction with respect to
symmetry $G_{\mathrm{q}_{1}}\otimes G_{\mathrm{q}_{2}}$.},
and exact relations between the masses of these particles and coupling
constant $\kappa$ are known \cite{F2} .

Here we consider the relations between QFTs (\ref{aGml}) and SMs in different
regions of integers $m,l$ and $h$. The simplest region corresponds to $m>>l$. 
In the limit $m\rightarrow\infty$ our QFT can be written as%
\[
\mathcal{A}_{G}=\mathcal{A}_{G}\left(  WZW_{l}\right)  -\kappa_{1}\int\sum
J_{a}\left(  z\right)  \overline{J_{a}}(\overline{z})d^{2}x
\]
i.e. Wess-Zumino-Witten model at level $l$ perturbed by currents. For finite
$m>>l,$ QFT (\ref{aGml}) describes the deformation of this SM. The scaling
function for the effective central charge on a circle of length $R$ was
calculated in \cite{F2} and is given by%
\begin{align*}
c_{\rm eff}(R)-c\left(  G,m,l\right)   &  =-2Dh\frac{l^{2}m^{3}}{h^{3}%
}(MR)^{4h/(m+l+h)}+\cdots
\qquad
m<<\log\left(  1/MR\right)
\\
&  =\frac{1}{2}Dl^{2}/\log^{3}(1/MR)+\cdots
\qquad m>>\log\left(  1/MR\right)
\end{align*}
in agreement with conformal perturbation theory and RG.

Another region depends on two parameters $m>>h,l>>h.$ It is similar to the
two-parameter deformations ($p_{1},p_{2}$) of principal chiral field (section
6). The QFT (\ref{aGml}) in this region describes the integrable deformations
of the principal chiral field considered as perturbed CFT. It is convenient to
introduce parameters $\textrm{k}=\frac{l-m}{l+m+2h}$ and $\textrm{y}=\frac
{h}{l+m+2h}<<1$. The scaling function for the effective central charge in this
region was calculated for $G=SU\left(  2\right)  $ and conjectured for all $G$
in \cite{F}. \ The result can be expressed through function $\tau(R)$ where%
\begin{equation}
\left(  \frac{1-\tau}{1+\tau}\right)  \left(  \frac{1+\mathrm{k}\tau
}{1-\mathrm{k}\tau}\right)  ^{\mathrm{k}}=\left(  MR\right)  ^{2\mathrm{y}%
}.\label{tauR}%
\end{equation}
Namely,%
\begin{equation}
c_{\rm eff}(R)-c\left(  G,m,l\right)  =-\frac{D\mathrm{y}}{\left(  1-\mathrm{k}^{2}\right)
2\tau}\left(  \left(  3+\mathrm{k}^{2}\right)  (1-\tau)^{2}-\mathrm{k}%
^{2}\left(  1-\tau^{2}\right)  ^{2}\right)  .\label{ctP}%
\end{equation}
This function has the asymptotics%
\begin{equation}
-D\mathrm{y}\left(  6\left(  \frac{1-\mathrm{k}}{1+\mathrm{k}}\right)  ^{2\mathrm{k}%
}\left(  MR\right)  ^{4\mathrm{y}}+\frac{64}{\left(  1-\mathrm{k}^{2}\right)
}\left(  \frac{1-\mathrm{k}}{1+\mathrm{k}}\right)  ^{3\mathrm{k}}\left(
MR\right)  ^{6\mathrm{y}}+\cdots\right)  \label{tpuV}%
\end{equation}
for $\mathrm{y}\log\left(  \frac{1}{MR}\right)  >>1.$ The first two terms in
this expansion were calculated exactly in \cite{F2}. The result coincides
with (\ref{tpuV}) up to $\mathrm{y\log y}$.
 
In the limit $\mathrm{y}%
\log\left(  \frac{1}{MR}\right)  <<1,$ corresponding to the principal chiral
field we get%
\begin{equation}
c_{\rm eff}(R)-D=-\frac{3D}{2\log(\frac{1}{MR})}+O\left(  \mathrm{y}\log\left(
\frac{1}{MR}\right)  \right)  \label{PCM}%
\end{equation}
in agreement with RG for principal chiral field.\footnote{The UV asymptotics (\ref{PCM}) in the region $\mathrm{y\log}\left(  \frac
{1}{MR}\right)  <<1$ (small deformations) is\ universal and for
perturbed CFT (\ref{aGml}) and SM (\ref{clim}) it is the same. On the other hand, for these theories the UV asymptotics in the
region $\mathrm{y\log}\left(  \frac{1}{MR}\right)  >>1$ are
different (see \cite{FOZ},\cite{F} and compare eqs. (\ref{cRRR}) and (\ref{ceffR})). This difference is related with the fact
that the quantum group restriction cancels the contribution of zero modes, which
provide the main term in the asymptotics of non-restricted SMs.}

It is interesting also to consider the perturbed $\mathcal{M}\left(
G,m,l\right)  $ CFTs in the region $h>>1,$ $m,l,$ fixed. We consider here
$G=SU\left(  k\right)  ,O\left(  k\right)  $ ($h=k,k-2$). The central charge
(\ref{cG}) will be now $2ml+O\left(  \frac{1}{h}\right)  $ and $ml+O\left(
\frac{1}{h}\right)  $ for $SU\left(  k\right)  $ and $O\left(  k\right)
.$\ The perturbation field $\Phi_{f}$ now will be associated with the
fundamental (vector) representation $f$ of $G.$

For $G=SU(k)$ and $m=1,$ $\Phi_{f}=\Lambda_{f}$ \ and in general case we use
as the perturbation field $\Phi\left(  \theta\right)  =e^{\mathrm{i}\theta
/k}\Phi_{f}+e^{-\mathrm{i}\theta/k}\Phi_{f}^{\ast}$ with dimension
$1-\Delta_{f}=\frac{1+km}{2(k+m)}+\frac{1+kl}{2(k+l)}=\frac{m+l}{k}+O\left(
\frac{1}{k^{2}}\right)  .$ Here $\theta$ is a topological parameter. \ This
QFT will have the \textquotedblleft instanton\textquotedblright\ and
perturbative contributions and in the limit $k\rightarrow\infty$ it will
coincide with $\frac{U(l+m)}{U\left(  l\right)  U\left(  m\right)  }$ SM with
the topological parameter. This QFT is K\"{a}hler SM and possesses the
instantons. It is natural to assume that it will be massive for $\theta=0$ and
massless for $\theta=\pi$ and if we make topological parameter $\theta$ to be
fluctuating field $\theta(x)$ (axion), the corresponding QFT will be integrable.

For $G=O(k)$ and $m=1$ the CFT perturbed by $\Phi_{f}$\ was studied by Fendley
\cite{FE}. For general $m$ field $\Phi_{f}$ has the dimension $1-\Delta
_{f}=\frac{1}{2}(\frac{m-1}{m+k-2}+\frac{l-1}{l+k-2}).$ In the limit
$k\rightarrow\infty$ this QFT describes integrable $\frac{O(l+m)}{O\left(
l\right)  O\left(  m\right)  }$ sigma models.

\section{Concluding Remarks}

In this section we provide some notes which can useful for studying dual
representations of deformed SMs on symmetric spaces.

\bigskip

$\mathit{1.}$ \ In this paper we did not considered the massless RG flows which
have the same UV asymptotics as the deformed principle chiral SMs and IR
asymptotics corresponding to rational CFTs. These Ricci flows are related with
SMs with singular metrics and will be studied in a separate publication. In
the previous section we have shown that integrable perturbed
rational coset models $\mathcal{M}(  G,m,l)  $ (\ref{aGml}) with
$m,l\gg h$ and $\kappa>0$ describe the massive deformed principle chiral SMs.
The massless RG flows in perturbed coset models $\mathcal{M}(G,m,l)$ with
$m,l\gg h,m\gg l$ are described by the action (\ref{aGml}) with $\kappa<0$.
The effective central charge $c_{\rm eff}(R)$ for these flows is
defined by eq. (\ref{ctP}) where one should make the substitution $\tau\rightarrow\tau_{-},$ where
$\tau_{-}(R)$ is defined by the relation %
\[
\left(  \frac{1-\tau_{-}}{1+\tau_{-}}\right)  \left(  \frac{1+\mathrm{k}%
\tau_{-}}{1-\mathrm{k}\tau_{-}}\right)  ^{\mathrm{k}}=-\left(  MR\right)
^{2\mathrm{y}}.
\]
It has the following UV and IR asymptotics:%
\[
c_{\rm eff}(R)-c(G,m,l)=-D\mathrm{y}\left(  6\left(  \frac{1-\mathrm{k}}{1+\mathrm{k}%
}\right)  ^{2\mathrm{k}}\left(  MR\right)  ^{4\mathrm{y}}-\frac{64}{\left(
1-\mathrm{k}^{2}\right)  }\left(  \frac{1-\mathrm{k}}{1+\mathrm{k}}\right)
^{3\mathrm{k}}\left(  MR\right)  ^{6\mathrm{y}}+\cdots\right)
\]
for $\left(  MR\right)  \ll1$ and
\[
c_{\rm eff}(R)-c(G,m-l,l)=\frac{D\mathrm{y}}{\mathrm{k}}\left(  6\left(  \frac{1-\mathrm{k}}{1+\mathrm{k}%
}\right)  ^{\frac{2}{\mathrm{k}}}\left(  MR\right)  ^{\frac{4\mathrm{y}%
}{\mathrm{k}}}-\frac{64}{\left(  1-\mathrm{k}^{2}\right)  }\left(
\frac{1-\mathrm{k}}{1+\mathrm{k}}\right)  ^{\frac{3}{\mathrm{k}}}\left(
MR\right)  ^{\frac{6\mathrm{y}}{\mathrm{k}}}+\cdots\right)
\]
for $\left(  MR\right)  \gg1$, i.e. these RG flows relate CFTs $\mathcal{M}%
\left(  G,m,l\right)  $ and $\mathcal{M}\left(  G,m-l,l\right)  .$

\bigskip

$\mathit{2.}$ \ In the Introduction we noted that SMs with
nice duality form a very small subspace in the space of
integrable SMs on the deformed symmetric spaces. Here we give a simple
example of the integrable QFT, which possesses a duality, but not nice duality. This example is provided by
$N=2$ supersymmetric Sine-Gordon model. Its QFT representation fits the two-parameter family
of integrable theories (\ref{2p}) where one of the parameters is fixed as $a_{1}^{2}=2$.
Similarly, its SM representation (\ref{tpfam}) is described by the two-parameter family with $p_{2}=2.$
This theory tends to $N=2$ supersymmetric $O(3)  $ (or $CP(1)  $) in the limit $a_{1}\rightarrow\infty,$ but it does not have the weak
coupling region. However, as it was noted in section 6, even in this case the
QFT representation  (\ref{2p}) is useful because it helps to reconstruct the
$S$-matrix. As can be seen from equations given in (\ref{int2p}), the currents $J_{2}^{(\pm)}$
have the spin $\frac{3}{2}$ and generate $N=2$ supersymmetry. The currents
$J_{1}^{(\pm)}$ generate $SU(2) _{\mathrm{q}}$ symmetry with
$\mathrm{q}=\exp\left(\mathrm{i}\frac{2\pi}{a_{1}^{2}}\right)=\exp\left(\mathrm{i}\frac{2\pi}{p_{1}}\right)$ that fixes $S$-matrix (\ref{SSp}).

\bigskip

$\mathit{3.}$ \ All known \textquotedblleft nice\textquotedblright\ dual theories
do not have the UV divergences in the perturbation theory. In particular, the
bulk vacuum energy is UV finite (see (\ref{sum})). This implies the following
equality %
\begin{equation}%
{\textstyle\sum\limits_{i=1}^{n_{b}}}
M_{b,i}^{2}=%
{\textstyle\sum\limits_{i=1}^{n_{f}}}
M_{f,i}^{2},\label{cfen}%
\end{equation}
where $M_{b,i}^{2}$ and $M_{f,i}^{2}$ are the masses of bosonic and Majorana
fermionic particles~in the weak coupling region $b\ll1$. This condition ensures the vanishing
of the one-loop UV divergence in the bulk vacuum energy
\[
\frac{1}{2}%
{\textstyle\sum\limits_{i=1}^{n_{f}}}
M_{f,i}^{2}\log L-\frac{1}{2}%
{\textstyle\sum\limits_{i=1}^{n_{b}}}
M_{b,i}^{2}\log L.
\]

The theory is integrable. This implies the existence of an integrable purely bosonic
QFT which is obtained by setting all fermions to zero in the weak coupling region.
The simplest examples of such QFT are provided by affine Toda theories, for
which the l.h.s. of (\ref{cfen}) is known. In the strong coupling (SM) region
we expect that some bosonic particles disappear from the spectrum, whereas the stable
bosonic particles and Majorana fermions have the same mass $M$ and form the
multiplet of the group in the numerator of the coset. The dimension of such
multiplet equals $D_{m}=$ $n_{f}+n_{b}^{(s)},$where $n_{b}^{(s)}$ is the
number of stable bosons. The UV central charge of such QFT is given by
$c_{UV}=n_{b}+\frac{1}{2}n_{f}.$ These equalities together with
(\ref{cfen}) put rather rigid restrictions on the \textquotedblleft
nice\textquotedblright\ dual theories with affine Toda bosonic part in the weak
coupling region.\footnote{If we assume that the unstable particles in the weak coupling region are the
bound states of fermionic particles and stable particles have the same mass
$M$ as all fermionic particles (see \cite{FL}) we derive for affine Toda QFTs
with the affine Lie algebras $\mathfrak{g}_{n}$ of rank (or number of bosonic
particles) $n=n_{b}$ the following results. Namely, $\mathfrak{b}_{n}:$ $%
{\textstyle\sum}
m_{b}^{2}=(2n-1)M^{2},\ n_{b}^{(s)}=2;\ \mathfrak{d}_{n}:$ $%
{\textstyle\sum}
m_{b}^{2}=(2n-2)M^{2},\ n_{b}^{(s)}=2;\ \mathfrak{c}_{n,dual}:$ $%
{\textstyle\sum}
m_{b}^{2}=M^{2},\ n_{b}^{(s)}=0$. The cases $\mathfrak{g}_{n}=\mathfrak{b}%
_{n,dual},\mathfrak{c}_{n},\mathfrak{a}_{n}$ were considered in \cite{FL} and
in this paper.} It is easy to see that for the theories considered in this
paper and in \cite{FL} they are satisfied.

\bigskip

$\mathit{4.} \ $Here we consider conditions for the existence of an infinite family
of integrable QFTs with \textquotedblleft nice\textquotedblright\ duality
without assumptions about purely bosonic part and UV convergence in
perturbation theory. It means that such QFTs can be described by weak coupling
theories in one regime and by slightly deformed SMs on the symmetric spaces in
the dual regime. Of course, when we know scattering theory of these QFTs for all
values of coupling parameters, the necessary condition can be easily
formulated. There should be a point in the space of parameters where the
scattering matrix is the identity. If we consider the asymptotically free SMs on
the coset spaces with one group $G$ in the numerator, the scattering matrix
should have the symmetry of this group $G$ (or $G$ $\times G$ in the case of
the principle chiral field) and the particles with minimal mass should belong
to some representation $R_{\min}$ of Yangian of $G$ (or $G$ $\times G$ for
principal chiral field) with dimension $D_{\min}$. Let $N_{\min}=D_{\min}$ for
real $R_{\min}$ and $N_{\min}=2D_{\min}$ for complex $R_{\min}.$\ For QFTs
with \textquotedblleft nice\textquotedblright\ duality the particles in the
weak coupling region are described by a set of almost free fields in the
action. A part of these particles disappears from the spectrum during the
evolution from weak to strong coupled QFT. In any case, even if we assume that
all particles survive in the strong coupling regime and were described by
fermionic fields in the weak coupling regime, we get a rather rough inequality%
\[
c_{UV}\geq\frac{1}{2}N_{\min}%
\]
This inequality excludes all principle chiral field QFTs, besides $G=SU\left(
2\right)  $ and all reasonable coset SMs besides $\frac{O\left(  n\right)
}{O(n-1)}$ considered in \cite{FL}, $\frac{SU(n)}{SU(n-1)}$ considered in
this paper and $\frac{Sp\left(  n\right)  }{Sp\left(  n-1\right)  }$ which
will be considered elsewhere.

\section*{Acknowledgements}

The author is very grateful to Alexey Litvinov for useful discussions and Sergei Alexandrov
for kind help in preparation of this manuscript for publication.

\appendix

\section{Explicit expressions for the currents and densities of local
integrals}

\label{ap1}

In this Appendix we provide the expressions for the current $\mathcal{W}_{3}$
for CFT (\ref{ac1}) with $n=3$ and the local integrals of motion for QFTs
(\ref{asl}) with $n=3$. We choose the two dimensional vectors $h_{i}$ (the
weights of fundamental representation of $su(3)$)$,$ and one dimensional
$\eta_{i}$ (the weights of the fundamental representation of $su(2)$) in the
form:%
\[
h_{1}=\left(\frac{1}{\sqrt{6}},\frac{1}{\sqrt{2}}\right),
\quad
h_{2}=\left(\frac{-2}{\sqrt{6}},0\right),
\quad
h_{3}=\left(\frac{1}{\sqrt{6}},\frac{-1}{\sqrt{2}}\right),
\quad
\eta_{1}=-\eta_{2}=\frac{1}{\sqrt{2}}
\]
Then the current $\mathcal{W}_{3}(3,b)$ has the form%
\begin{align}
\mathcal{W}_{3}  &  =\sqrt{6}b\partial\varphi_{1}\left(  bc(\partial
\varphi_{1})^{2}+\mathrm{i}\sqrt{6}b^{2}\partial\varphi_{1}\partial
\vartheta_{0}-3cb(\partial\varphi_{2})^{2}+\sqrt{2}ca(a\partial^{2}\varphi
_{2}+2\mathrm{i}b\partial^{2}\vartheta_{3})\right) \nonumber\\
&  +3\mathrm{i}b\partial\vartheta_{0}\left(  2b^{2}(\partial\varphi_{2}%
)^{2}+a(3a(\partial\vartheta_{3})^{2}+(3+5b^{2})(\partial\vartheta_{0}%
)^{2}-\sqrt{2}b\left(  2b\mathrm{i}\partial^{2}\vartheta_{3}+a\partial_{2}%
^{2}\partial^{2}\varphi_{2}\right)  \right) \nonumber\\
&  +9\sqrt{2}ab\partial^{2}\vartheta_{0}(b\partial\vartheta_{3}-\mathrm{i}%
a\partial\varphi_{2})-3\sqrt{6}c\partial^{3}\varphi_{1}, \label{w3k}%
\end{align}
where $a=\sqrt{1+b^{2}},c=\frac{1}{\sqrt{6}}\sqrt{3+b^{2}}.$ The current
$\mathcal{W}_{3}(3,k)$ can be derived from (\ref{w3k}) by substitution
$b^{2}=-3-k.$

The local integral of motion for QFT (\ref{asl}) with $n=3$ is
\begin{align*}
\mathcal{I}_{3}^{(3)}  &  =\partial\varphi_{1}\left(  \sqrt{2}b((\partial
\varphi_{1})^{2}-3(\partial\varphi_{2})^{2})-6(a^{2}\partial^{2}\varphi
_{2}+\mathrm{i}ab(\sqrt{3}\partial^{2}\phi_{1}+\partial^{2}\phi_{2})\right) \\
&  +2\mathrm{i}\partial\phi_{0}\left(  3b^{2}((\partial\varphi_{1}%
)^{2}+(\partial\varphi_{2})^{2})+3a^{2}((\partial\phi_{1})^{2}+(\partial
\phi_{2})^{2})+(b^{2}+a^{2})(\partial\phi_{0})^{2}\right) \\
&  -\mathrm{i}\sqrt{2}a\partial\phi_{1}((\partial\phi_{1})^{2}-3(\partial
\phi_{2})^{2})-6\mathrm{i}ab\partial\varphi_{2}\left(  \partial^{2}\phi
_{1}-\sqrt{3}\partial^{2}\phi_{2}\right)  -6a^{2}\partial\phi_{1}\partial
^{2}\phi_{2}%
\end{align*}
here $a=\sqrt{1+b^{2}}.$

\section{Possible parametrization of SU$\left(  n\right)  $}

\label{ap2}

In this Appendix we describe the possible parametrization of matrix $g\subset
SU\left(  n\right)  .$ The generators of $SU\left(  n\right)  $ can be
numerated as $T_{ij}^{\left(  +\right)  },T_{ij}^{\left(  -\right)  }$ with
$n\leq$ $j>i\geq1$ and $H_{i},i=1,..,n-1$ where $H_{i}$ are the real diagonal
matrices $tr(H_{i}H_{j})=2\delta_{ij}.$ Let $E_{i,j}(k,m)=\delta_{ik}%
\delta_{jm}$ then%
\[
T_{ij}^{\left(  +\right)  }=E_{ij}+E_{ji},
\qquad
T_{ij}^{\left(  -\right)
}=\mathrm{i}(E_{ij}-E_{ji})
\]
The matrix $g\subset SU\left(  n\right)  $ can be written as%
\[
g=%
{\displaystyle\prod\limits_{k=1}^{n-1}}
e^{\mathrm{i}\eta_{k}H_{k}}%
{\displaystyle\prod\limits_{j>i}^{n}}
e^{\mathrm{i}(a_{ij}T_{ij}^{\left(  +\right)  }+b_{ij}T_{ij}^{\left(
-\right)  })}%
\]
The multipliers in this product can be easily calculated, for example, for
$n=2,T_{ij}^{\left(  +\right)  }=\tau_{1},T_{ij}^{\left(  -\right)  }=\tau
_{2}$ \ and%
\[
g=e^{\mathrm{i}\eta_{1}\tau_{3}}\left(
\begin{array}
[c]{cc}%
{\small \cos\theta}_{12} & {\small \mathrm{i}e}^{-\mathrm{i}{\small \beta
}_{12}}{\small \sin\theta}_{12}\\
{\small -\mathrm{i}e}^{\mathrm{i}{\small \beta}_{12}}{\small \sin\theta} &
{\small \cos\theta}_{12}%
\end{array}
\right)=e^{\mathrm{i}\eta_{1}\tau_{3}}\mathrm{g}_{12}({\small \theta}_{12},\beta_{12})
\]
where $\theta_{12}=\sqrt{a_{12}^{2}+b_{12}^{2}}$ and $e^{\mathrm{i}\beta_{12}%
}=\frac{a_{12}+\mathrm{i}b_{12}}{a_{12}-\mathrm{i}b_{12}}.$ It is easy to
generalize this expression to any multiplier $e^{\mathrm{i}(a_{ij}%
T_{ij}^{\left(  +\right)  }+b_{ij}T_{ij}^{\left(  -\right)  })}.$ We denote
$\mathrm{g}_{ij}({\small \theta}_{ij},\beta_{ij})=e^{\mathrm{i}(a_{ij}%
T_{ij}^{\left(  +\right)  }+b_{ij}T_{ij}^{\left(  -\right)  })}$ then
$g\subset SU\left(  n\right)  $ can be written as%
\[
g=%
{\displaystyle\prod\limits_{k=1}^{n-1}}
e^{\mathrm{i}\eta_{k}H_{k}}%
{\displaystyle\prod\limits_{j>i}^{n}}
\mathrm{g}_{ij}({\small \theta}_{ij},\beta_{ij})
\]
We note that order of multipliers in this product can be taken arbitrary. It
leads to different parametrizations of $g.$

\bigskip

\end{document}